\documentclass[iop,apj]{emulateapj}
\usepackage{mathrsfs,amsmath,amssymb,amstext}
\usepackage{natbib}
\usepackage{epsfig}
\usepackage{color}
\usepackage{float}
\usepackage{graphicx}
\usepackage{gensymb}
\slugcomment{}

\shorttitle{Isophotal Structure}
\shortauthors{B.C. Ciambur}

\begin{document}

\newcommand{\elli}{$\textsc{Ellipse}$}
\newcommand{\iso}{$\textsc{Isophote}$}
\newcommand{\bmo}{$\textsc{Bmodel}$}
\newcommand{\cmo}{$\textsc{Cmodel}$}
\newcommand{\ifit}{$\textsc{Isofit}$}

\title{Beyond { Ellipse}(s): Accurately Modelling the Isophotal Structure \\of Galaxies with {\it Isofit} and {\it Cmodel}}
\author{B. C. ~Ciambur}
\affil{Centre for Astrophysics and Supercomputing, Swinburne University of Technology, Hawthorn, VIC 3122, Australia}
\email{bciambur@swin.edu.au}

\begin{abstract}

This work introduces a new fitting formalism for isophotes which enables more accurate modelling of galaxies with non-elliptical shapes, such as disk galaxies viewed edge-on or galaxies with X-shaped/peanut bulges. Within this scheme, the angular parameter which defines quasi-elliptical isophotes is transformed from the commonly used, but inappropriate, polar co-ordinate to the `eccentric anomaly'. This provides a superior description of deviations from ellipticity, better capturing the true isophotal shape. Furthermore, this makes it possible to accurately recover both the surface brightness profile, using the correct azimuthally-averaged isophote, and the two-dimensional model of any galaxy: the hitherto ubiquitous, but artificial, cross-like features in residual images are completely removed. The formalism has been implemented into the \textsc{IRAF} tasks \elli\;and \bmo\;to create the new tasks `\ifit', and `\cmo'. The new tools are demonstrated here with application to five galaxies, chosen to be representative case-studies for several areas where this technique makes it possible to gain new scientific insight. Specifically: properly quantifying boxy/disky isophotes via the fourth harmonic order in edge-on galaxies, quantifying X-shaped/peanut bulges, higher-order Fourier moments for modelling bars in disks, and complex isophote shapes. Higher order ($n > 4$) harmonics now become meaningful and may correlate with structural properties, as boxyness/diskyness is known to do. This work also illustrates how the accurate construction, and subtraction, of a model from a galaxy image facilitates the identification and recovery of over-lapping sources such as globular clusters and the optical counterparts of X-ray sources.\\
\end{abstract}


\maketitle

\section{Introduction}\label{sec:Introduction}

Galaxies come in a wide variety of shapes and sizes. Among the numerous efforts throughout the years to impose some order among their ever growing numbers, one particular classification scheme remains persistent to this day: grouping galaxies by their shape as it is seen at optical and infrared wavelengths. This is essentially a structural classification -- while some galaxies, known as `late type' on the Hubble/Jeans sequence (\citealt {Jeans1919}, \citealt{Hubble1926}, \citealt{Jeans1928}), display disks, spiral arms, bars and bulges, others are more relaxed systems, and are commonly referred to as `early type', elliptical or lenticular galaxies. However, even `elliptical' galaxies are rarely truly elliptical. Their isophote shapes often deviate from pure ellipses, in a characteristic way. These deviations, the most common being referred to as `boxyness' or `diskyness', originate from the structure of the stellar orbits that make up the galaxy. Because of the physical link between isophote shape and galaxy (structural) properties, quantifying these deviations provides a valuable tool to study galaxies. Numerous works have revealed correlations between boxyness/diskyness and physical properties such as kinematics (\citealt{Carter1978}, \citealt{Davies1983} \citealt{Lauer1985}, \citealt{Carter1987}, \citealt{Bender1988a}, \citealt{Peletier1990}, \citealt{Jaffe1994}), brightness profiles (\citealt{Nieto1991}) and even global radio and X-ray properties (\citealt{Bender1989}). 

However, the formalism used in the past (and to this day) to describe boxyness/diskyness in isophotes has limited applicability. In particular, it is only efficient when applied to galaxy isophotes which are relatively well described by pure ellipses and the deviations from ellipticity are small (a few per cent). This has limited isophotal structure studies to only a subsample of the whole galaxy population which met the above conditions, specifically elliptical or early type galaxies. Even in such objects, the presence of e.g. embedded disks caused the models to fail and produce residual maps marked by ubiquitous crosses or artificial features. The literature is literally full of such examples (e.g., \citealt{Crosses2004}, \citealt{Duncan2}, \citealt{Duncan1}, \citealt{Crosses2010}, \citealt{Crosses2011}, \citealt{Crosses2011_2} \citealt{Crosses2013}, \citealt{Guerou_etal2015}, etc). Moreover, some of the correlations (or lack thereof) have been affected by the use of a formalism which fails to adequately capture the true isophotal shapes.

In this work a new isophote fitting formalism is introduced, which is capable of modelling galaxies with significantly more complex or exotic isophote structures, with a particular focus on disk galaxies viewed edge-on. The paper is structured as follows. Section \ref{sec:Iso-Fit} provides a short description of how deviations from perfect ellipticity are commonly expressed and modelled in the popular isophote analysis tool \elli. In Section \ref{sec:NewAngle} a new angular metric is introduced, which better expresses isophote shapes in general and is particularly powerful in modelling disk galaxies viewed edge-on, as well as galaxies with X-shaped/peanut bulges. The formalism is implemented and further demonstrated on a representative such galaxy. Section \ref{sec:App} demonstrates the applicability of this formalism to any isophote shape, and explores scientific case-studies where this new technique can provide insight. In this section, four additional galaxies are modelled and the possibility to quantify the peanut/X-shape bulge of some galaxies from their photometric structure is briefly explored. Finally, in Section \ref{sec:sum} the paper reiterates its main conclusions and proposes further potential scientific applications of the method.\\

\section{Isophote Fitting}\label{sec:Iso-Fit}

Among the early efforts to express departures from pure elliptical shapes in mathematical form, the work of \cite{Carter1978} was to become key in modelling more refined and realistic galaxy isophotes. This work proposed adding perturbations to an ellipse as a function of azimuthal angle $\phi$, in the same way as in a Fourier series decomposition 

\begin{equation}
	I(\phi) = \left <I_{ell}\right > + \sum_{n} \left[ A_{n} \textrm{sin}(n\phi) + B_{n} \textrm{cos}(n\phi) \right] ,
	\label{equ:FHarm}
\end{equation}

where $I(\phi)$ is the intensity profile along the isophote, expressed as a function of (central) azimuthal angle, $\left <I_{ell}\right >$ is the average intensity of the purely elliptical path, and the sum represents Fourier harmonic perturbations to $\left <I_{ell}\right >$, with $n$ being the harmonic (integer) order. Note that perturbing the \textit{intensity distribution} in this way is equivalent to distorting the \textit{physical shape} of the isophote. The two cases are used interchangeably throughout, for illustrative purposes.

\begin{figure}
 \centering
	\includegraphics[width=1.\columnwidth]{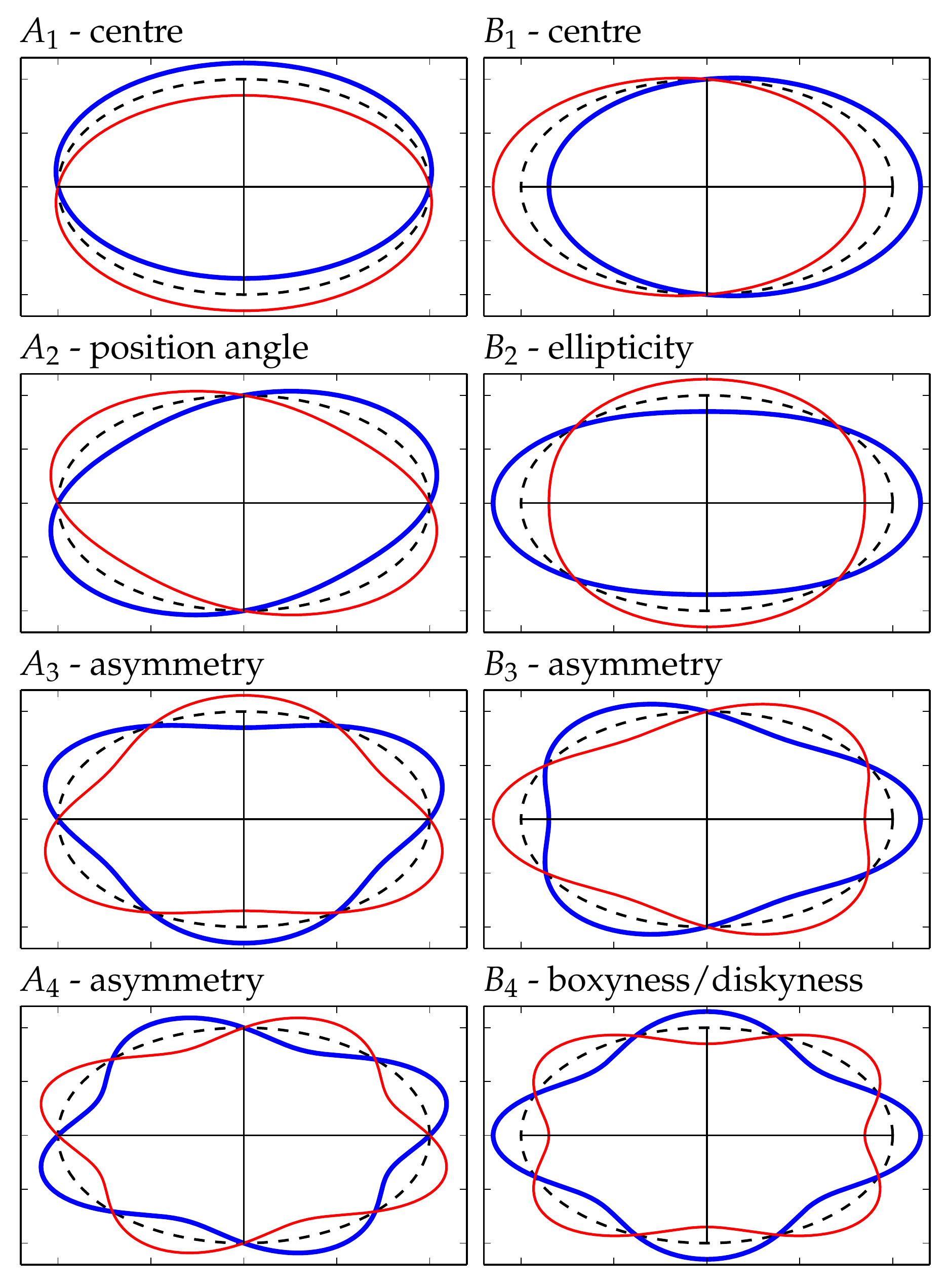}
	\caption{The physical significance of the first four harmonic corrections to an elliptical isophote. Positive coefficients are illustrated in blue (thick) while negative coefficients in red (thin). The reference isophote is plotted in black (dotted) in each panel.}
	\label{fig:FHarms}
\end{figure}

This formalism is particularly elegant as the coefficients of the Fourier harmonics ($A_{n}$ and $B_{n}$) carry physical meaning. This is illustrated in Figure \ref{fig:FHarms} for the first four harmonic orders. Here we observe how the harmonics bring corrections to the ellipse's centre ($n=1$), the ellipticity and position angle ($n=2$), as well as capturing various types of asymmetries and the boxy or disky feature (the $B_{4}$ coefficient).

These principles were later detailed into a full isophote fitting algorithm in the seminal paper of \cite{Jedrzejewski1987_paper} (also in \citealt{Jedrzejewski1987_proc}). The reader is referred to these works for a detailed description of the algorithm, which is only briefly outlined below. In this method, nested isophotes (not necessarily concentric) are fitted at several pre-defined points along the semi-major axis (denoted by $a$) of a galaxy in a CCD image. At each of these points, the isophote starts as a pure ellipse defined by initial guess-values for its geometric parameters, namely centre position $(x_{0},y_{0})$, position angle (PA) and ellipticity ($e = 1- b/a$, where $b$ is the semi-minor axis). The image is then sampled along this elliptical path, giving the 1-dimensional intensity distribution as a function of azimuthal angle $I_{data}(\phi)$, which is first modelled by the right-hand side of Equation \ref{equ:FHarm}, but restricted to $n \in \left\{ 1,2 \right\}$, i.e.,

\begin{equation}
	I(\phi) = \left <I_{ell}\right > + \sum_{n=1}^{2} \left[ A_{n} \textrm{sin}(n\phi) + B_{n} \textrm{cos}(n\phi) \right] .
	\label{equ:FHarm-1st2}
\end{equation}

After a least-squares minimisation of the quantity $S$ in Equation \ref{equ:resid_sq}, the best-fit values of the harmonic coefficients are used to update the geometric parameters of the ellipse, and then $\left <I_{ell}\right >$. 

\begin{equation}
	S = \sum_{i} \; \left[I_{data}(\phi_{i}) - I(\phi_{i})\right]^{2} .
	\label{equ:resid_sq}
\end{equation}

The minimisation iterates until a minimum root mean square criterion is met. By only considering the first two harmonics (centre, PA and $e$), this section of the algorithm essentially computes the best-fiting pure ellipse on the data isophote corresponding to this semi-major axis radius. Finally, higher order harmonic perturbations to the ellipse ($n \ge 3$) are fitted (sequentially) through the same iterative minimisation. \\

A powerful and robust implementation of this algorithm is in the \textsc{IRAF} (Image Reduction and Analysis Facility\footnote{http://iraf.noao.edu/}) external software package \iso, developed by the Space Telescope Science Data Analysis System\footnote{http://www.stsci.edu/institute/software\_hardware/stsdas} (STSDAS). The main tasks in this package (frequently referred to in the remainder of this work) are the isophote fitting task \elli, and the task \bmo\;which builds a 2-dimensional galaxy model based on the parameters obtained by \elli. This software has had considerable success since its release, performing admirably when modelling the type of galaxies it was designed to work with, which are objcts characterised in general by relatively low ellipticity ($e \lesssim 0.5$) or, for higher $e$, with isophotes very close to elliptical (low harmonic amplitudes). These are typically early-type galaxies with no (or with rather face-on) disks. The \elli\;task also performs quite well when modelling (late-type) disk galaxies viewed face-on, after the appropriate masking of biasing features such as dust lanes. 

However, for galaxies having more complex shapes or, in particular, for disk galaxies viewed at relatively high inclination or edge-on, \elli\;(and \bmo) begin to break down and give rise to residual images marked by ubiquitous cross-like patterns or characteristic, alternating regions of excess and deficit light (see Sections \ref{sec:NA-TestCase} and \ref{sec:App}). 

Obviously it is desirable to correctly model a galaxy's light distribution. A good model provides physical insight into the galaxy itself, but moreover, subtracting an accurate model from an image gives the possibility to perform meaningful studies of substructures still remaining after the subtraction, such as star clusters, globular clusters or optical counterparts to X-ray/radio sources. In the case of edge-on galaxies, a model where the light intensity is significantly underestimated along the major axis of the galaxy introduces errors in the surface brightness profile, which is typically measured (and in fact output by \elli) along $a$, and is one of the main quantities of interest provided by isophote fitting programs. The surface brightness profile is essential in galaxy decomposition and is routinely used to quantify galaxy components and characterise structural scaling relations (see \citealt{Graham2013_review} review article and references therein).\\

\section{A Natural Angular Metric}\label{sec:NewAngle}

\begin{figure}
 \centering
	\includegraphics[width=1.\columnwidth]{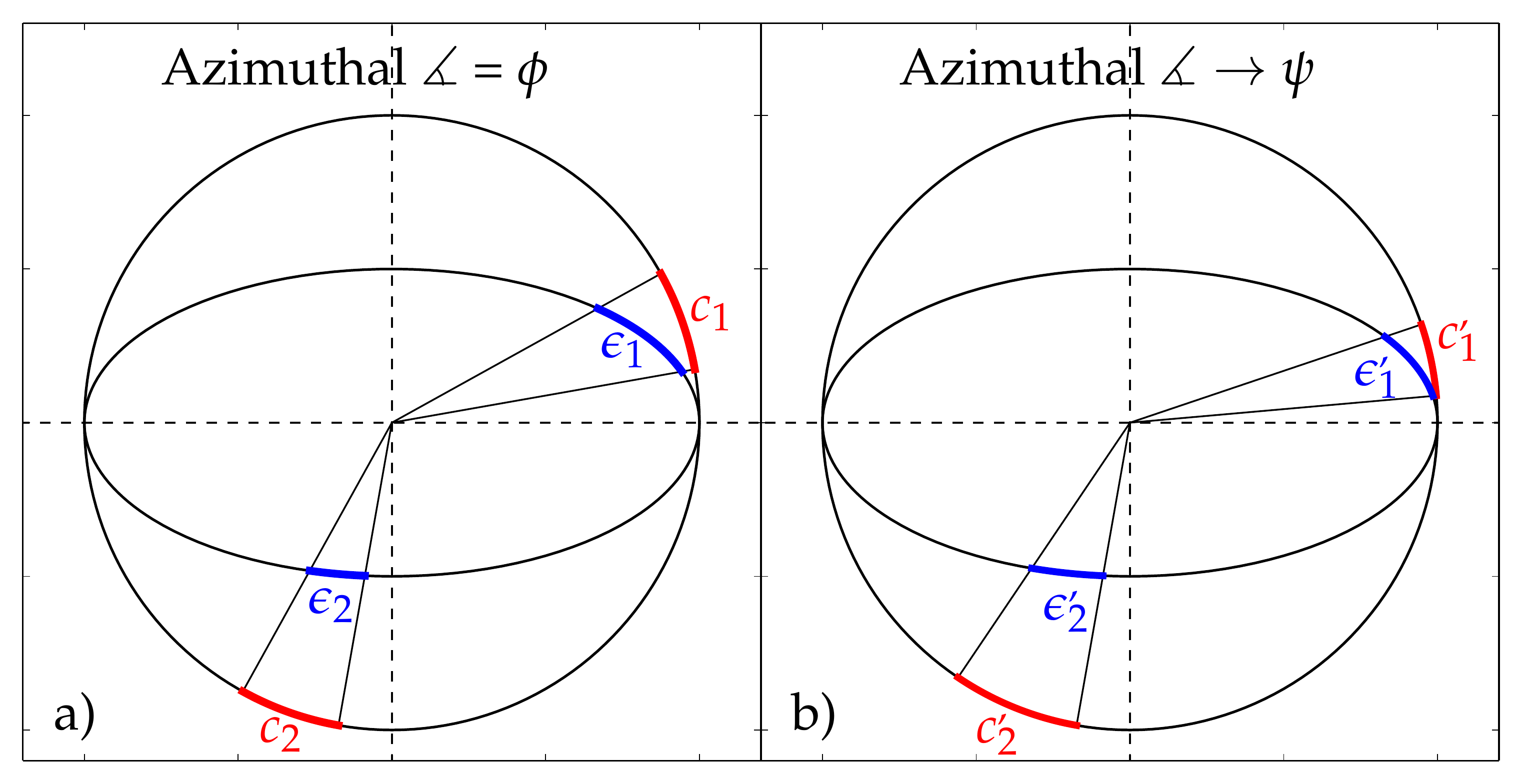}
	\caption{\textit{Panel} a): Equal polar opening angles ($\phi$) define equal arc lengths on a circle ($c_1 = c_2$) but not on an ellipse ($\epsilon_1 \neq \epsilon_2$). \textit{Panel} b): When the same opening angles are transformed to the `eccentric anomaly' ($\psi$), they change as a function of $\phi$ such that they define equal arc lengths on an ellipse ($\epsilon_1' = \epsilon_2'$), but no longer on the circle ($c_1' \neq c_2'$).  $\psi$ therefore samples an ellipse uniformly.}
	\label{fig:arcs}
\end{figure}

The reason why isophote-fitting algorithms like \elli\;fail in the regime of high $e$ and high harmonic amplitudes is traced back to the incorrect manner in which a quasi-elliptical isophote path is parameterised and sampled. The task \elli\;splits the isophote into sectors as it would a circle: it dividies the azimuthal range ($\phi \in [0;2\pi]$) uniformly, in equal bins. While the sectors (arc lengths) corresponding to equal steps in $\phi$ are all equal for a circle, they are not for an ellipse, but are actually longest along the major axis and shortest along the minor axis (see Figure \ref{fig:arcs}, left panel). This causes the isophote to be too coarsely sampled along $a$ (in the case of edge-on galaxies, $a$ corresponds to the plane of the disk, and thus needs to be very finely sampled as it contains most of the light), which then leads to bright streaks in residual images. \\

\subsection{The Eccentric Anomaly}\label{sec:NA-Theory}

This work introduces a new isophote fitting formalism implemented in a new \textsc{IRAF} task \ifit, where the Fourier harmonics that quantify deviations from perfect ellipses are expressed as a function of an angular parameter more natural to elliptical shapes, namely the `eccentric anomaly', henceforth denoted by $\psi$. The eccentric anomaly is frequently used in celestial mechanics to describe the position of a point orbiting on an elliptical path, in canonical form (i.e where the angle is defined from the orbit centre, not from one focus), or more generally to express the parametric equation of an ellipse. It is the `natural' angular co-ordinate for ellipses, and the $\psi$ co-ordinate of a point on an ellipse is in fact related to its azimuthal/plane-polar angular co-ordinate (denoted as $\phi$ throughout this paper) through the ellipticity $e \equiv 1 - b/a$, such that 

\begin{equation}
	\psi = -\mathrm{arctan}\left(\frac{\mathrm{tan}(\phi)}{1-e}\right) .	
	\label{equ:conversion}
\end{equation}

By transforming from $\phi$ to $\psi$, the isophotes are sampled uniformly along the entire azimuthal range (Figure \ref{fig:arcs}, right-hand panel), and the Fourier corrections are expressed as

\begin{equation}
	I(\psi) = \left <I_{ell}\right > + \sum_{n} \left[ A_{n} \textrm{sin}(n\psi) + B_{n} \textrm{cos}(n\psi) \right] .
	\label{equ:FHarmpsi}
\end{equation}

\begin{figure}
	\centering
	\includegraphics[width=1.\columnwidth]{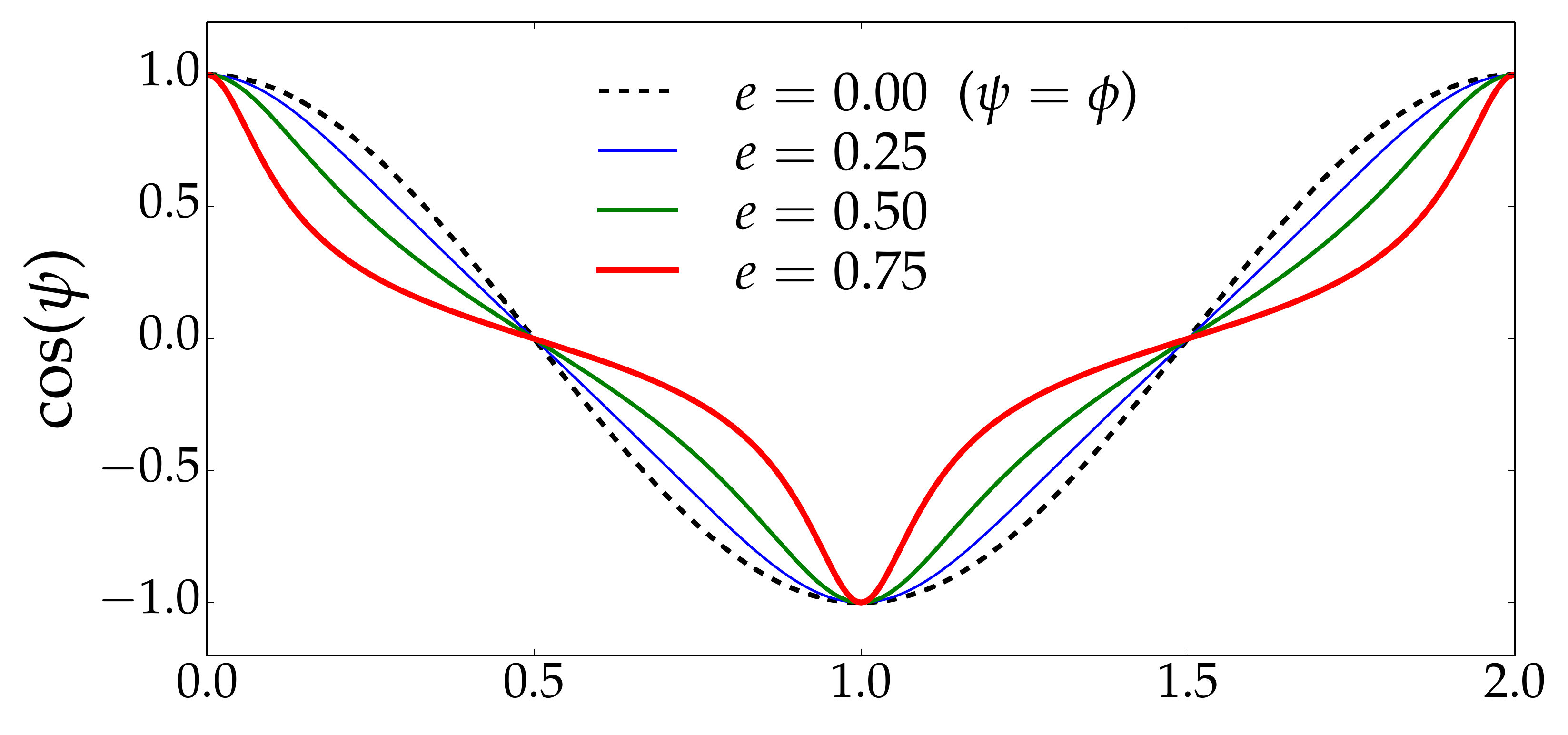}\\
	\includegraphics[width=1.\columnwidth]{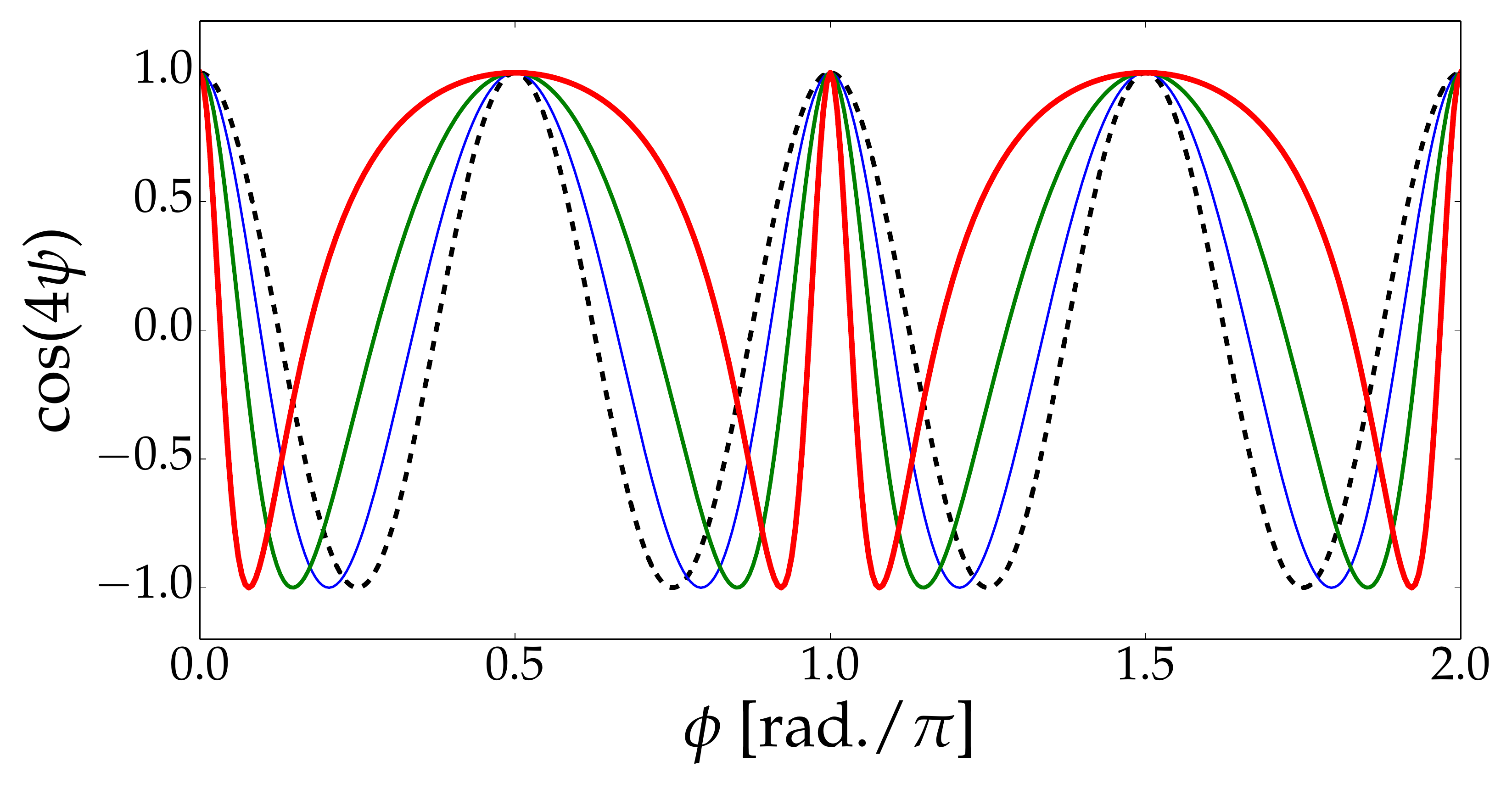}
	\caption{\textit{Top:} the cosine of the `eccentric anomaly' $\psi$, for different values of ellipticity ($e$) plotted against the polar angle $\phi$. The amplitude of this harmonic ($B_1$ in Figure \ref{fig:FHarms}) controls the centre position of the isophote. \textit{Bottom:} The cosine of $4\psi$, for the same values of $e$ as in the top panel. The amplitude of this harmonic ($B_4$ in figure \ref{fig:FHarms}) controls the boxyness/diskyness of the isophote.}
	\label{fig:cosine}
\end{figure}

Furthermore, the Fourier harmonics correct the isophotal shape quite differently. This is illustrated in Figure \ref{fig:cosine}, where the top panel shows how the cosine form of $\psi$ (which enters in Equation \ref{equ:FHarmpsi}) changes as the ellipticity changes from $e = 0$ ($\psi = \phi$) to high ellipticity. The bottom panel illustrates how the cosine part of the $n = 4$ harmonic (i.e., boxyness/diskyness) acts sharply at the major axis (defined to correspond to $\psi = 0,2\pi$ and $\psi = \pi$) and more softly along the minor axis ($\psi = \pi/2$ and $\psi = 3\pi/2$), as the ellipticity increases. The more flattened (i.e., high $e$) an isophote, the sharper the harmonic correction around $a$ and the softer the correction around $b$. This is precisely what is required, since a disk galaxy viewed edge-on has a very bright but thin disk dominating the light along $a$, while along and around the minor axis $b$ the light is dominated by a spheroidal bulge. The lower panel in Figure \ref{fig:cosine} thus illustrates how the `lemon-shaped' contours in the lower-right panel of Figure \ref{fig:FHarms} can be made more `pointy' along the major axis.

A concrete example of harmonic terms characterising an isophote shape is given in Figure \ref{fig:coef246}, which also highlights the difference between using the generic polar angle (\elli) and eccentric anomaly (\ifit) to express the Fourier corrections. Here we notice how the angle transformation ($\phi \rightarrow \psi$) makes the $A_2$ correction correspond more to a rotation (change in position angle), and the $B_4$ correction resemble more a disky, lemon- or diamond-shaped contour. The lower two panels show a higher ($n=6$) harmonic, which may carry physical meaning in edge-on galaxies with so-called X-shaped or `peanut' bulges. This tentative point is further explored in Section \ref{sec:App1_xpbulges} by modelling such a galaxy. \\

\begin{figure}
	\centering
	\includegraphics[width=1.\columnwidth]{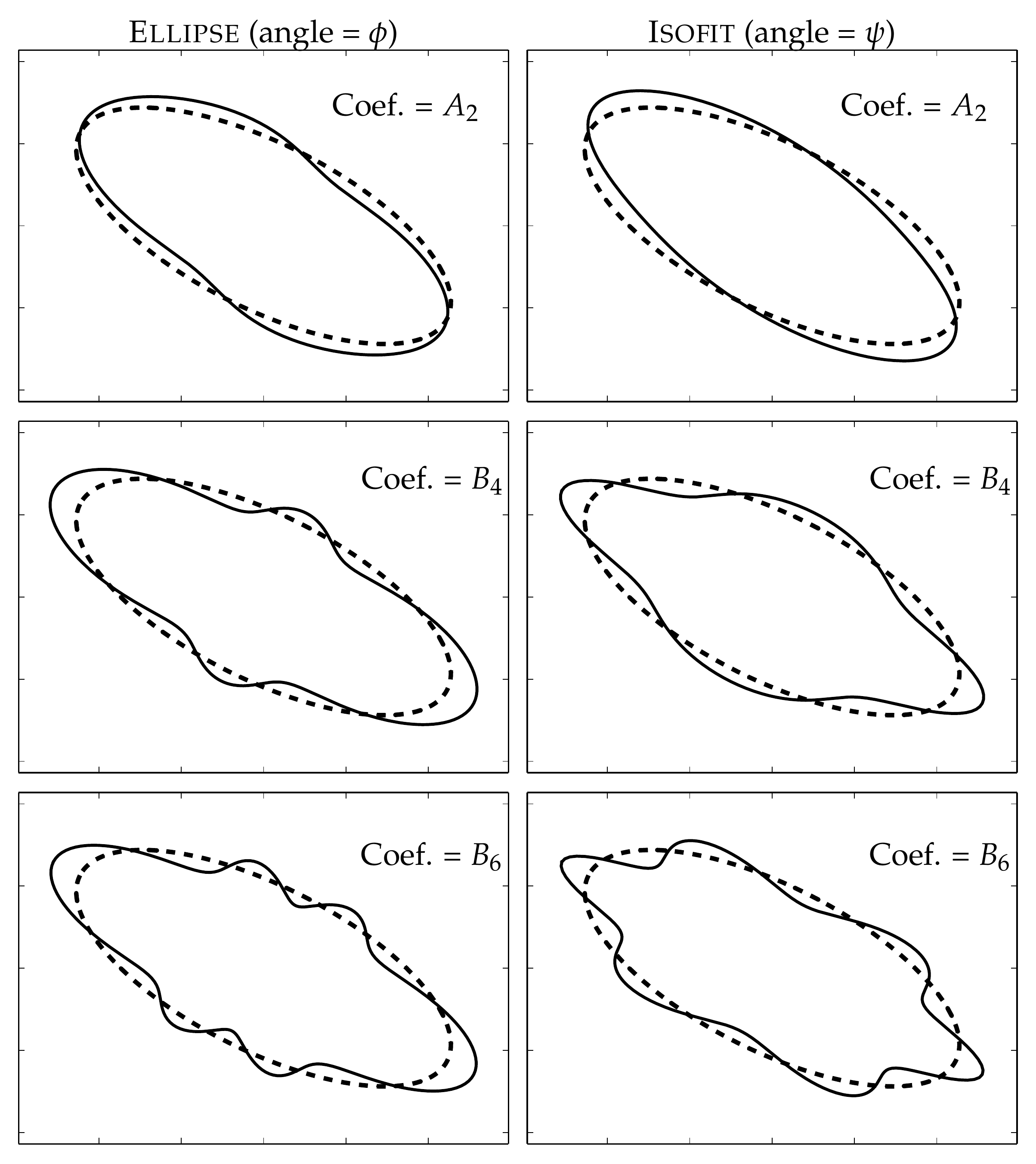}
	\caption{Examples of three harmonic corrections to an elliptical isophote, expressed as a function of polar angle $\phi$ (left panels) and in the new formalism (as a function of $\psi$; right panels), for values of $e=0.65$ and $A_2 = B_4 = B_6 = 0.15$, chosen to be relatively high to clearly illustrate the effect. In every panel, the reference ellipse is displayed with a dashed line.}
	\label{fig:coef246}
\end{figure}

\subsection{Implementation in \textsc{IRAF}}\label{sec:NA-Imp}

In order to model galaxy isophotes according to the formalism discussed in the previous section, the source code of the \textsc{IRAF} package \iso\; was modified at the harmonic fitting level (\elli) and 2D galaxy model building (\bmo). An angular parameter similar to the eccentric anomaly has been used in other galaxy-modelling codes, most notably in {\sc Galfit} \citep{Peng_etal2010}, where it takes a slightly different form (the Fourier formalism distorts the co-ordinate grid rather than the intensity distribution) and the authors consider only the cosine of the angle plus a phase in expressing the harmonics. 

The \elli\;task was modified and renamed \ifit\;in order to differentiate the new formalism from the standard \elli.  When fitting for the harmonic coefficients, \ifit\;now defines the isophotes as a function of the eccentric anomaly $\psi$ by transforming the azimuthal angle $\phi$ according to Equation \ref{equ:conversion} and applying the appropriate sign correction to each angle quadrant. Additionally, all harmonic corrections\footnote{In practce, $n \sim 8-12$ is the useful limit} of $n \geq 2$ are now added \textit{simultaneously}, in contrast to \elli, which fitted $n > 2$ harmonics sequentially. By fitting the harmonics simultaneously, all orders contribute together to construct the isophote (just as in a Fourier decomposition), rather than each individual order being a `best-attempt' at capturing the dominant components of the galaxy light distribution. For example, in the particular case when the dominant component is a bright, thin disk along the major axis (i.e., an edge-on galaxy), we have seen in Figure \ref{fig:FHarms} and in Figure \ref{fig:coef246} (bottom two panels), and one can also deduce from Equation \ref{equ:FHarm}, that every even-$n$ $B_n$ harmonic provides a correction at the positions corresponding to $a$ ($\psi = 0$ and $\psi=\pi$). Therefore, by fitting them sequentially each coefficient is optimised to best capture the disk feature rather than all orders contributing together to model less obvious features over the whole azimuthal range.

In \elli, the harmonics are chosen by the user and their number is technically unlimited. This same flexibility in terms of choice is kept in \ifit, though the simultaneous fitting does limit the number of useful harmonics. Nevertheless, in practice the harmonic amplitude decreases with increasing $n$ and usually asymptotes to zero for $n \sim 10-12$ and above, even for extremely thin edge-on disks. Furthermore, in the majority of cases, odd-$n$ harmonics bring little-to-no refinement to the isophote shapes (unless the galaxy has significant asymmetries), so by restricting the choice to only even orders, one can reliably fit harmonics up to $n = 12$ and obtain accurate models, as is demonstrated in Section \ref{sec:NA-TestCase} below. 

In addition to \elli, the auxilliary task \bmo\; (`build model') was modified and renamed \cmo\; (`construct model'). The task \cmo\; considers all harmonics included by the user (\bmo\; only used $n \leq 4$) and generates a 2D model of the galaxy based on the fitting results output by \ifit. The harmonic corrections are applied as a function of $\psi$, which is obtained as before. In the regime $e \rightarrow 0$, the standard \elli\;result is recovered, since in this limit $\psi = \phi$.

\subsection{Detailed Example in Practice: the Edge-on Galaxy ESO 243-49}\label{sec:NA-TestCase}

The ideal test-cases, which best highlight the improvements brought by the formalism discussed in this work, are edge-on disk galaxies. These galaxies are characterised by diamond-like or lemon-like isophote shapes which deviate strongly from pure ellipses. They are the objects for which the standard version of \elli\;gives the least accurate results in both 1D (systematically underestimated major-axis surface brightness profiles) and 2D (cross-like features in residual images). 

The S0 galaxy ESO 243-49 was chosen as the test-case for the new tasks \ifit\;and \cmo. This edge-on galaxy has been intensly studied following the discovery of a source of hyperluminous X-ray emission  (referred to as HLX-1) associated with a potential intermediate-mass black hole within the galaxy but outside the nuclear region (\citealt{Farrell_etal2009}, \citealt{Webb_etal2010}). The fact that HLX-1 was found to have an optical counterpart \citep{Soria_etal2010} makes this galaxy particularly interesting to model, as one of the aims of this work is to achieve high-quality 2D galaxy models and, after subtracting them from the image data, to perform accurate photometry on the remaining substructure (such as HLX-1) left behind in the residual image. Additionally, ESO243-49 also displays a relatively large number of star clusters both in the plane of the disk and outside, as well as a dust disk in the nucleus, also viewed edge-on. The image data for ESO~243-49 was retrieved from the Hubble Legacy Archive\footnote{http://hla.stsci.edu} and consisted of an $H$-band image taken with the Wide Field Camera 3 (WFC3, IR channel, F160W filter, PI: S. A. Farrell) on the {\it Hubble Space Telescope} (HST). 

To fully showcase the improvements brought by the new formalism, the image was modelled firstly with the standard \elli/\bmo\;and then with \ifit/\cmo. Both versions were run, where applicable, with identical initial conditions. \\

\subsubsection{Constructing the Isophotes}\label{sec:NA-TestCase-iso}

\begin{figure}
	\centering
	\includegraphics[width=.9\columnwidth]{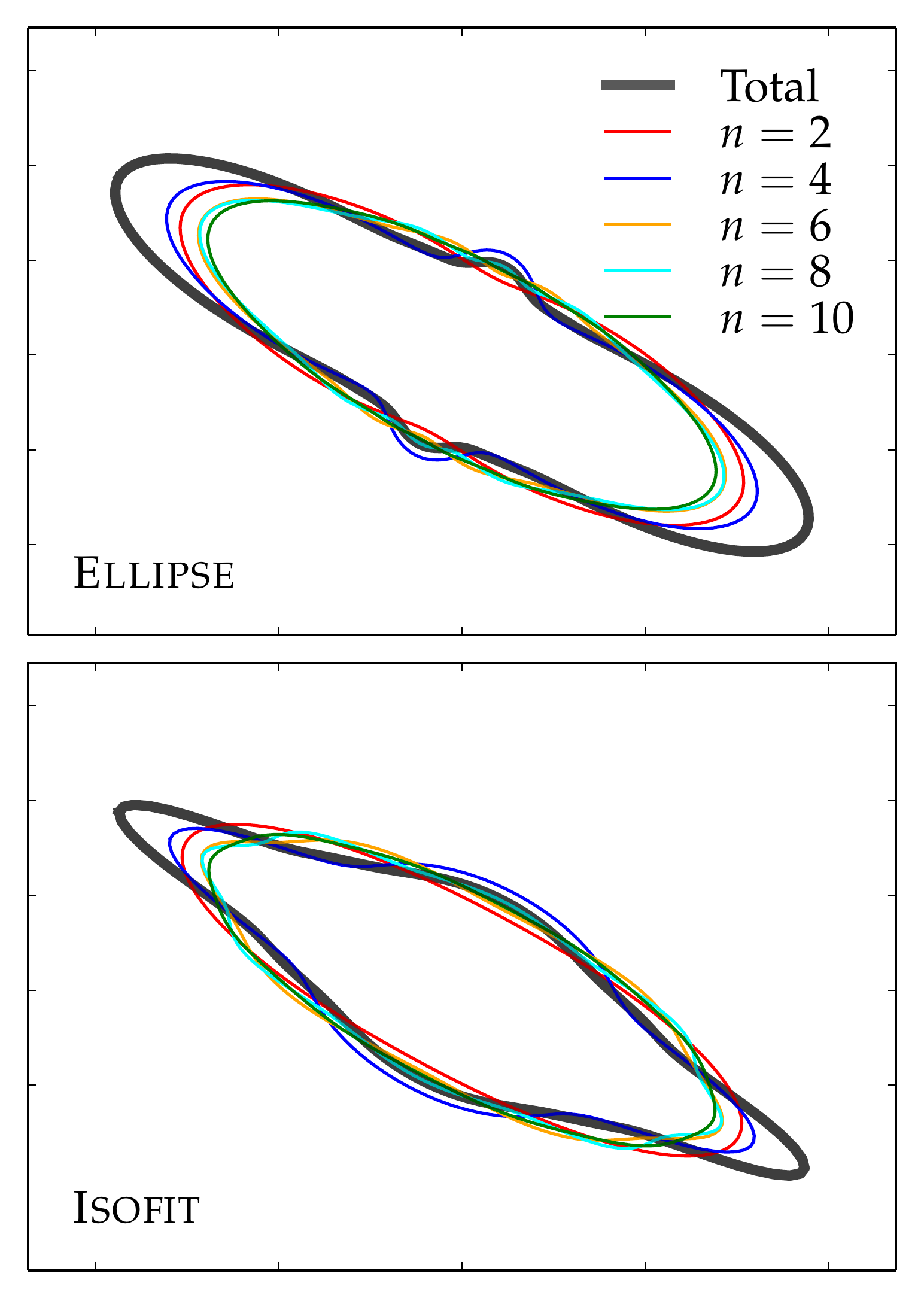}
	\caption{An example of how an isophote of ESO 243-49 (thick black line) is constructed from several harmonic orders, with the standard $I(\phi)$ formalism (top) and the new $I(\psi)$ formalism (bottom). The different harmonic orders, denoted by $n$, refer to Equation \ref{equ:FHarm} (top) and Equation \ref{equ:FHarmpsi} (bottom).}
	\label{fig:contributions}
\end{figure}

An important observation concerning the construction of each isophote is that, due to the use of the polar angle $\phi$ instead of $\psi$ to define the isophote in \elli, the high-order ($n \gtrsim$ 4) harmonics do very little to improve its shape (see Figure \ref{fig:coef246}, bottom-left panel). This is in part why, generally, higher order harmonics have not been used in the literature and it has been assumed that they are not important. This is illustrated in the top panel of Figure \ref{fig:contributions}, which shows a single isophote and its constituent \textit{even} Fourier contributions for 2 $\geq n \geq$ 10. This isophote was constructed from the harmonic coefficients computed by \elli\;while modelling ESO~243-49 The bottom panel in Figure \ref{fig:contributions} displays an isophote constructed at the same semi-major axis but with the new formalism of \ifit\;proposed in this work. Here we see how each upper harmonic order brings a new level of refinement to the final shape, as one expects in fact from a Fourier analysis. The ability of each of these corrections to describe the actual isophote shape (at this semi-major axis) in ESO~243-49 will become apparent in Section \ref{sec:NA-TestCase-2D}.

\subsubsection{1D Improvements}\label{sec:NA-TestCase-1D}

A very useful quantity routinely extracted from galaxy images with the use of isophote fitting tools is, of course, the surface brightness profile $\mu (R)$ -- the radial distribution of light in the galaxy. This quantity is frequently used when performing one-dimmensional structural decompositions (e.g., bulge/disk) of galaxies. Often times $\mu$ is measured along the major axis of the galaxy. For edge-on galaxies, this is precisely the locus where light is systematically underestimated when using improperly defined isophote shapes. In such cases (and the test-case ESO 243-49 is a prime example) one must resort to estimating $\mu (R)$ by taking a crude `cut' along the image major axis, which is undesirable for several reasons. Firstly, this method is noisy, because of e.g., dust, star clusters or foreground stars which happen to be located in the plane of the disk. Further, when taking a cut, one obtains a single measurement of the brightness at each radial position, whereas by fitting the isophote, one obtains a value for the brightness at a given $R$ which is the average value across the whole azimuthal range. Secondly, a cut is a less accurate representation of the major axis surface brightness profile because, in general, the position angle of the isophotes does not remain constant with increasing radius: real isophotes often rotate with increasing $R$, following features such as bars (see Section \ref{sec:substruct}), so a 1D cut capturing this effect must be a curve, not a line.

Figure \ref{fig:eso243-49_profile} shows how the use of the old formalism in \elli\; systematically underestimates the major axis surface brightness profile of the test-case galaxy ESO 243-49, compared with a (constant PA) direct cut taken along the disk plane and assumed to be a more realistic approximation of the true $\mu(R_{\rm maj})$. The discrepancy is most prominent across the range ($5 \lesssim R_{\rm maj} \lesssim 30$ arcsec) where the disk light dominates. The formalism presented in this work does remarkably better and is the closest and most accurate representation of $\mu(R_{\rm maj})$, for the reasons explained above. The discrepancy between the profiles computed with \elli\;and \ifit\;is maximal at $R_{\rm maj} = 16.5$ arcsec, where \elli\; underestimates $\mu$ by 0.77 mag arcsec$^{-2}$, which is a factor of two in surface brightness. 

Such a discrepancy propagates into the bulge/disk decompositions and leads to wrong results. Judging by the shape of the underestimated profile (green dashed curve in Figure \ref{fig:eso243-49_profile}), the disk component appears to follow an exponential form. If modelled as such, it would eat into and impact the fit of the central bulge (its brightness, concentration, effective radius etc.). When considering the correct profile (black solid curve in Figure \ref{fig:eso243-49_profile}), the disk component is in reality more appropriately described by an inclined disk model (a S\'ersic function with $n_{S\acute{e}rsic}  < 1$; \citealt{Pastrav2013}), which leads to quite a different decomposition result, with a brighter and more concentrated (higher S\'ersic index) bulge. When performing bulge/disk decompositions, it is common for the major-axis surface brightness profile, $\mu(R_{\rm maj})$, to be mapped onto the `equivalent' axis $R_{\rm eq}$. This mapping is equivalent to transforming each isophote of semi-major axis $R_{\rm maj}$ into a circle of equal enclosed area, which then has a radius $R_{\rm eq}$. This way, one can use azimuthal symmetry to compute \textit{integrated} quantities from $\mu (R_{\rm eq})$, including the magnitudes of the model components that are fit to this profile. This is fully consistent with the formalism implemented in this work as the Fourier harmonic perturbations change the shape of the isophote but conserve its area. This is demonstrated in the Appendix.

\begin{figure}
	\centering
	\includegraphics[width=1.\columnwidth]{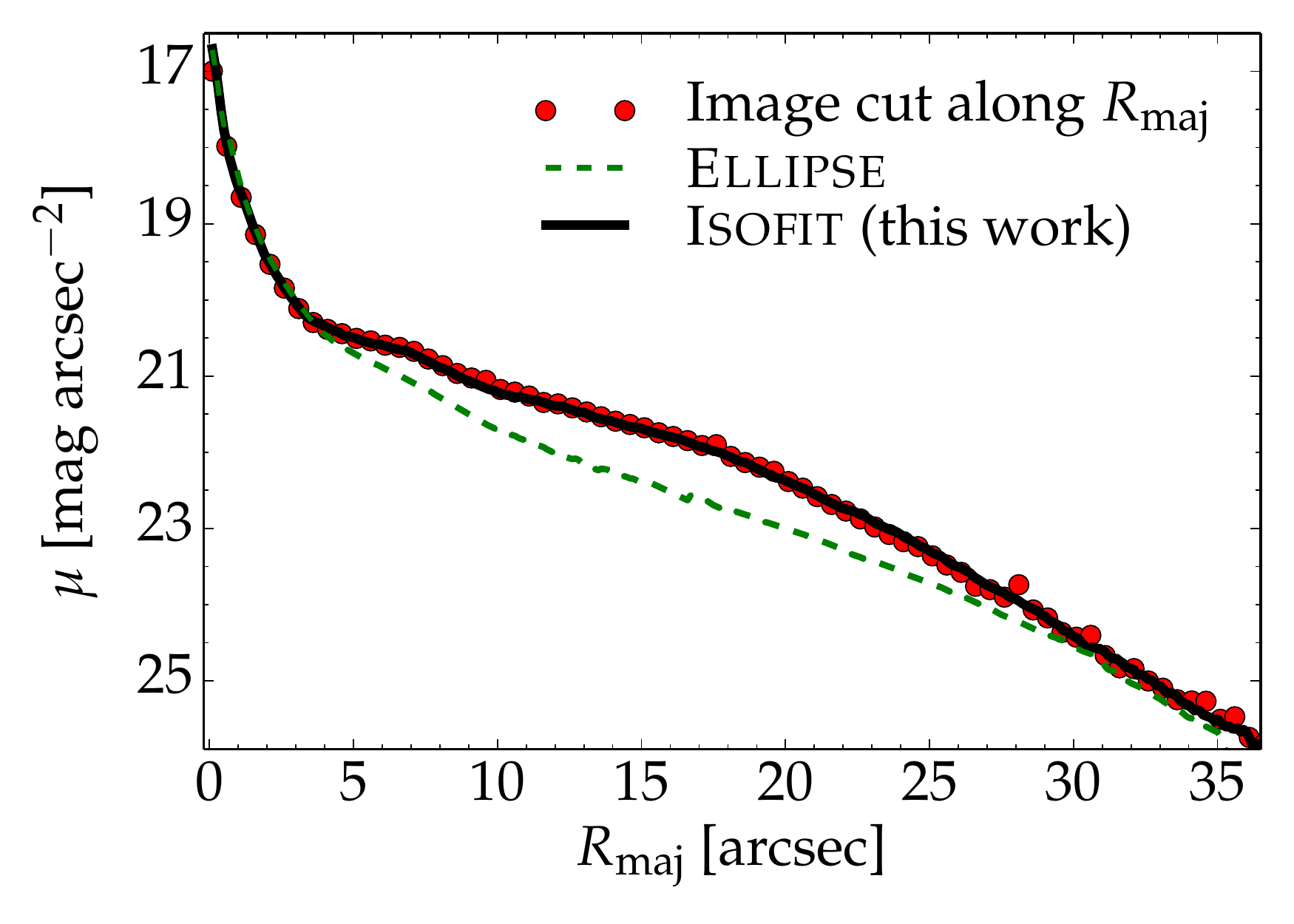}
	\caption{Major axis surface brightness profile of ESO 243-49. The standard \elli\;task systematically underestimates the profile in the range $5 - 30$ arcsec (where the disk dominates), by as much as a factor of 2 in surface brightness occurring at $R_{maj} = 16.5$ arcsec. See the lower panels of Figure \ref{fig:eso2543-49} to understand this.}
	\label{fig:eso243-49_profile}
\end{figure}

A final note on the 1D aspect of this work is that using the correct angular metric to express the isophotes makes it possible to obtain the surface brightness profile not just of the major axis, but also the minor axis or in fact any direction, directly by using $\left<I_{ell} \right>$ and the harmonic coefficients (all provided by \ifit) in Equation \ref{equ:FHarmpsi}, for any azimuthal angle. This was not possible before, and we can readily see it just by inspecting the single isophote shown in the upper panel of Figure \ref{fig:contributions} (the shape as a function of azimuthal angle is wrong).\\

\subsubsection{2D Improvements}\label{sec:NA-TestCase-2D}

\begin{figure*}
	\centering
	\includegraphics[width=1.\textwidth]{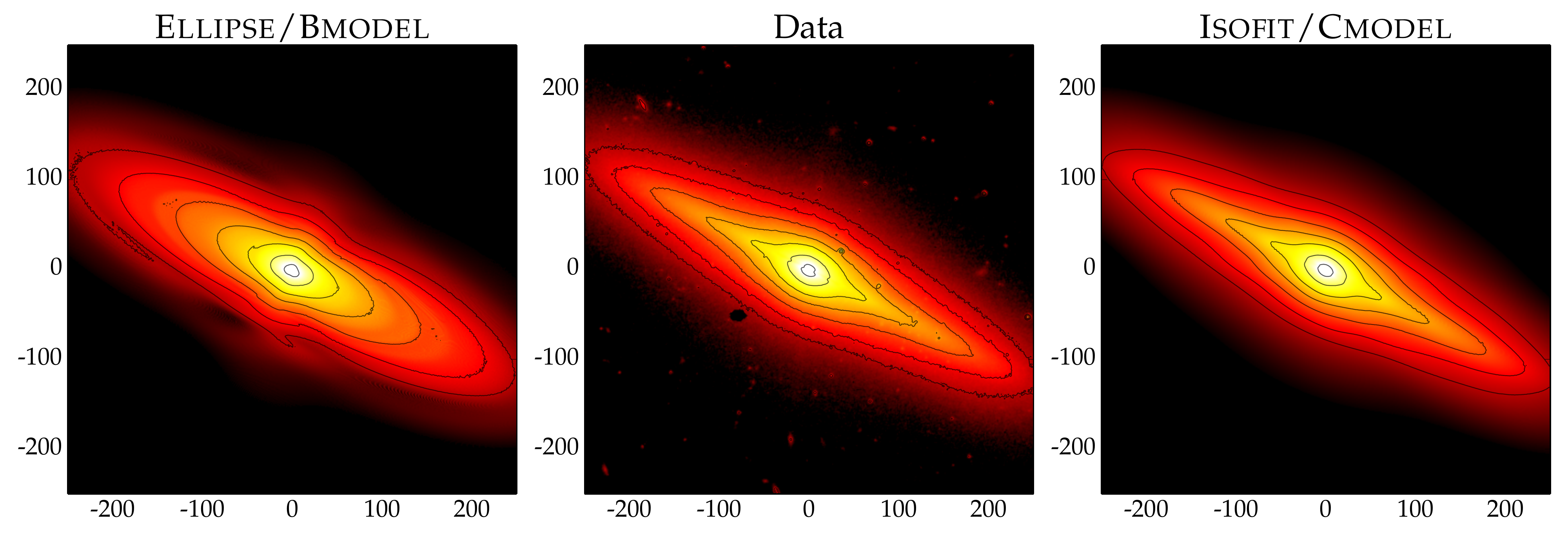}\\
	\flushright
	\includegraphics[width=.9\textwidth]{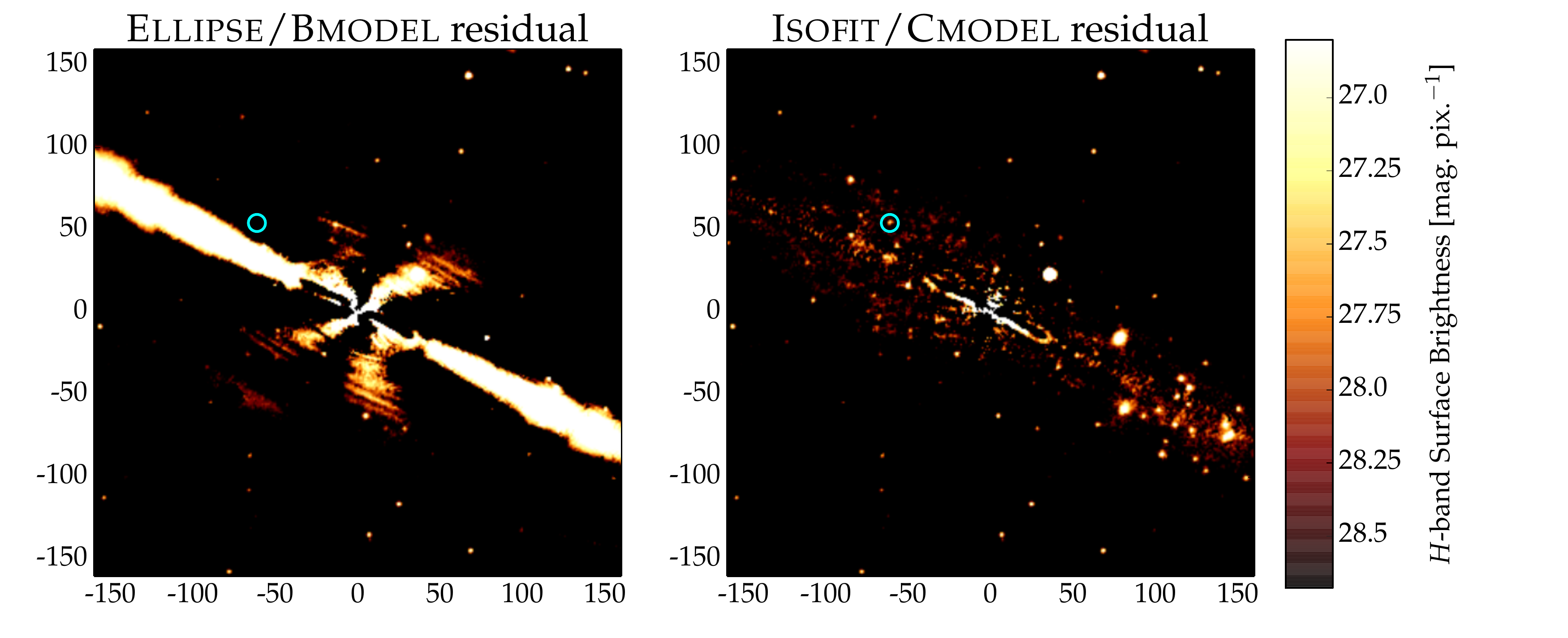}
	\caption{{\it Top 3 panels}: 2D model of the galaxy ESO 243-49 constructed with old formalism of \elli\;and \bmo\; (top-left); HST image of ESO 243-49 (top-centre); 2D model constructed with \ifit\; and \cmo\; (top-right). {\it Bottom 2 panels}: Residual images obtained after subtracting the \bmo-based galaxy model (bottom-left) and the new \cmo-based galaxy model (bottom-right)  from the galaxy image. The position of the optical counterpart of HLX-1 is marked with the cyan circle in both residual maps. The higher harmonic terms were turned on and used to construct both models. However, the old formalism used in \elli/\bmo\;fails due to the reasons detailed in Section \ref{sec:NewAngle}.}
	\label{fig:eso2543-49}
\end{figure*}

The auxilliary task \bmo\; in the \textsc{IRAF} package \iso\;has been routinely used in the past to generate two-dimensional models of galaxies, based on the 1D information computed by \elli. Due to the improper angular metric employed, the residuals obtained from subtracting these 2D models from galaxy images are notorious for displaying obvious artificial features, in particular for systems with strong departures from pure elliptical shapes. In this subsection \bmo\;and the new task \cmo, are both run on the test-case galaxy ESO 243-49. The input information for \bmo\; consisted of the best-fit generated by \elli, while the input for \cmo\;consisted of the best-fit generated by \ifit. The results are shown in Figure \ref{fig:eso2543-49}.

We can clearly recognise in the upper-left panel of Figure \ref{fig:eso2543-49} the wrong shape of a disky isophote resulting from improperly defined ellipses (we noticed this in Figures \ref{fig:coef246} and \ref{fig:contributions}). The inevitable consequence is that when such a model is subtracted from the galaxy image, the resulting residual map is contaminated by artificial features (for ESO 243-49, the residual image is dominated by such an artificial feature along the disk plane, which is quite obvious in the bottom-left panel of Figure \ref{fig:eso2543-49}).

It is immediately apparent that \cmo\; (top-right and bottom-right panels in Figure \ref{fig:eso2543-49}) achieves a remarkably better representation of the galaxy light distribution than the previous standard (\bmo).  With \cmo\; it is now possible to construct realistic and accurate 2D models of galaxies. Having such models can bring considerable insight into the structure and morphology of galaxies and also enables meaningful studies of the substructure left behind after subtracting the accurate model from the image. As to the latter, \cmo\; makes it possible to easily identify the optical counterpart of the X-ray source HLX-1 and compute its near-infrared brightness. This was done with the software Aperture Photometry Tool \citep{APT-Laher_etal2012}, resulting in the apparent $H$-band magnitude $m_H = 22.68 \pm 0.27$ (in the Vega magnitude system).\\

\section{New Science - Case Studies}\label{sec:App}

This section demonstrates the much broader applicability of the new technique to describing virtually any galaxy. Four additional galaxies are modelled, chosen to be representative of a specific class that presents scientific interest and is the object of active current research. The following is by no means an exhaustive study of each class, but it is meant to merely illustrate the usefulness of the method introduced in this work, and to highlight some tentative ways to investigate such galaxies and quantify their properties.

\subsection{Peanut/X-shaped Bulges \& Boxy+Disky Isophotes}\label{sec:App1_xpbulges}

\begin{figure*}
	\centering
	\includegraphics[width=1.\textwidth]{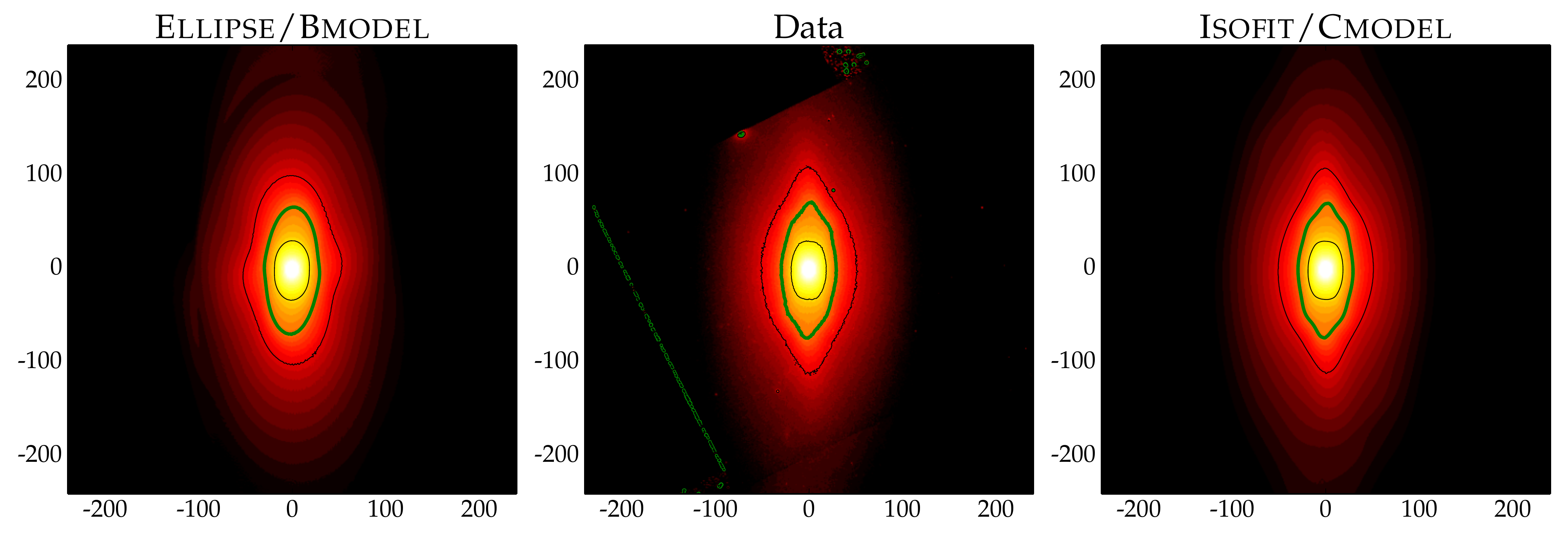}\\
	\flushright
	\includegraphics[width=.9\textwidth]{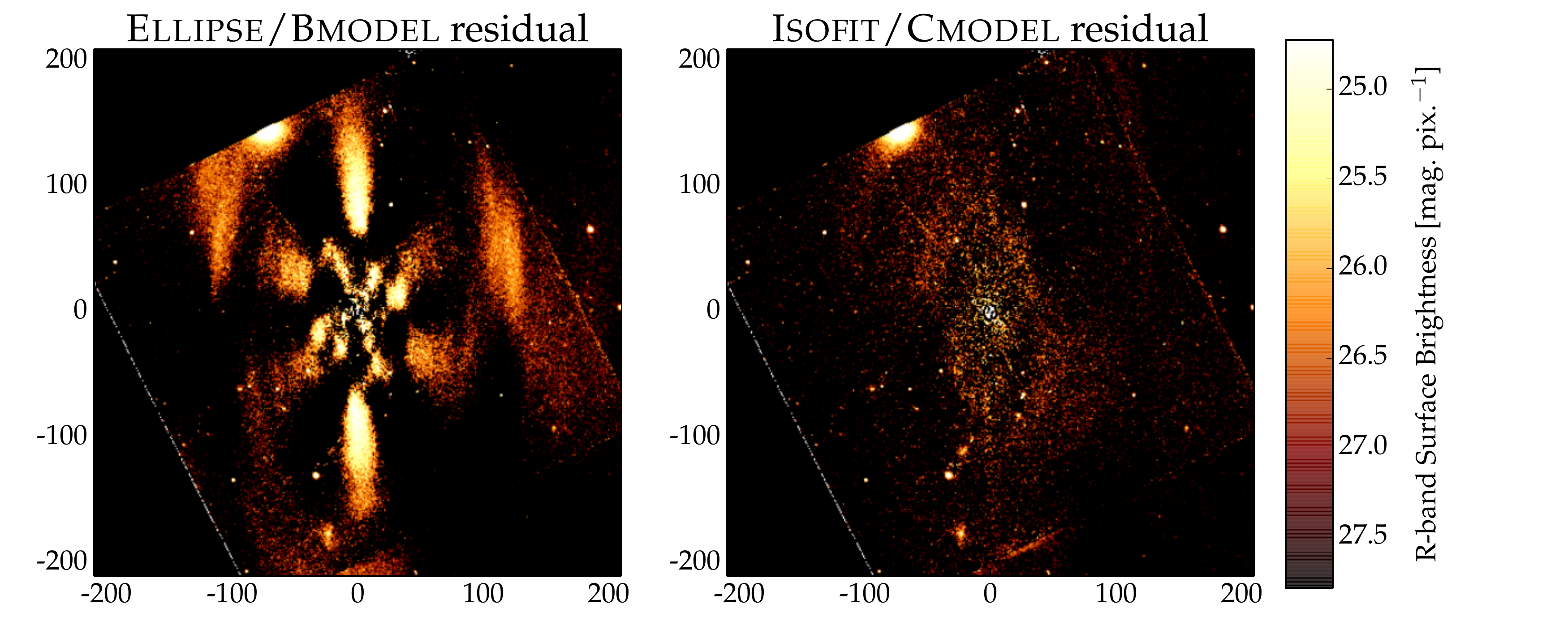}
	\caption{NGC 2549: a galaxy with a known X-shaped bulge. The panels are arranged analogously to Figure \ref{fig:eso2543-49}. The upper panels use a segmented colormap to better highlight the X-shaped bulge feature, and only 3 contours are overplotted: the inner and outer black contours have boxy and disky shapes, respectively. The middle contour (green) is a combination of both and needs higher orders to be properly modelled.}
	\label{fig:n2549}
\end{figure*}

The bulges of many nearby galaxies have been observed to display non-elliptical shapes, particularly obvious in (but not restricted to) galaxies where the bulge is embedded in a large scale disk. These bulges are commonly referred to as boxy, peanut-shaped or X-shaped, and have been amply studied both observationally and with simulations (e.g \citealt{deVaucouleurs1974},  \citealt{CombesSanders1981}, \citealt{Bureau_etal1999}, \citealt{Bureau_etal2006} -- see \citealt{Laurikainen2015} for a recent review). In terms of formation scenarios, it is believed that this structural feature has a kinematical origin, resulting from the re-organisation of stellar orbits within the bulge, and is possibly related to the formation of bars. Because peanut/X-shaped bulges are especially obvious when the galaxies are viewed edge-on, these objects make ideal candidates for the modelling technique described in this work.

A representative galaxy, known to have an X-shaped bulge, was chosen to be modellled in this section, namely NGC 2549. The imaging data was retrieved from the HLA, and consisted of an $R$-band HST image taken with the Wide Field and Planetary Camera 2 (WFPC2, filter F702W; PI: W. Jaffe). As before, the galaxy was modelled with the old method (standard \elli\;, \bmo) and the new technique (\ifit, \cmo), both run with identical starting conditions. The results are displayed in Figure \ref{fig:n2549}.

\begin{figure*}
	\centering
	\includegraphics[width=1.\textwidth]{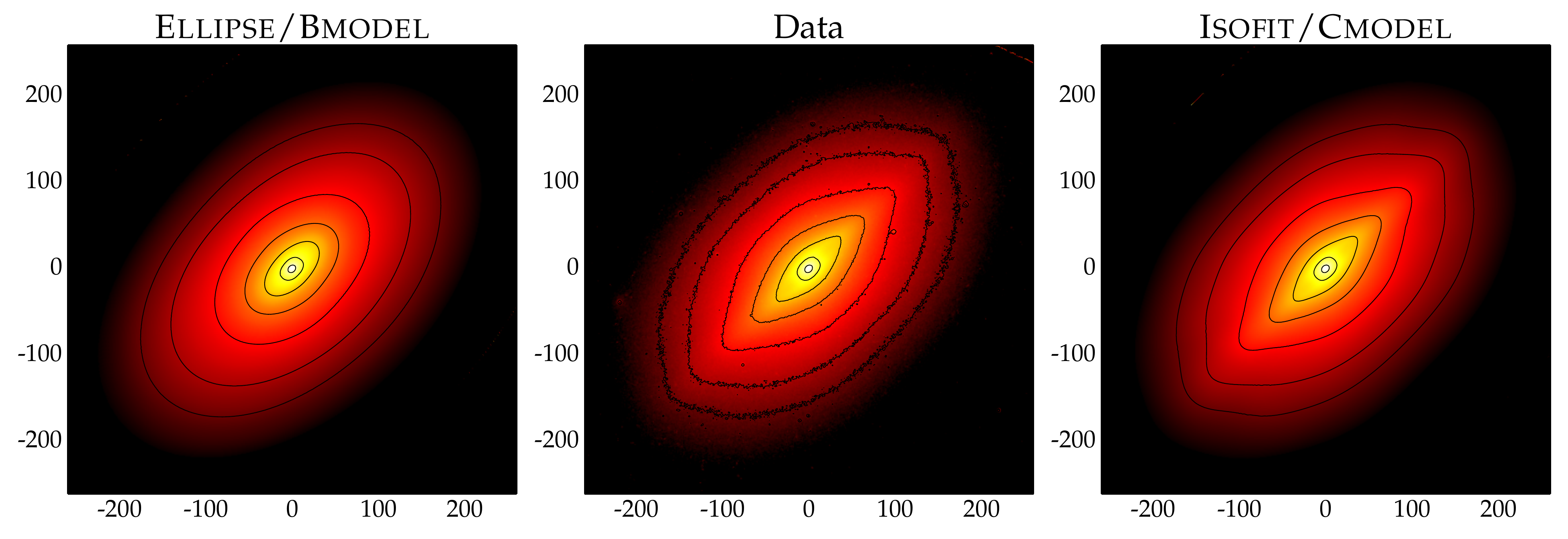}\\
	\flushright
	\includegraphics[width=.9\textwidth]{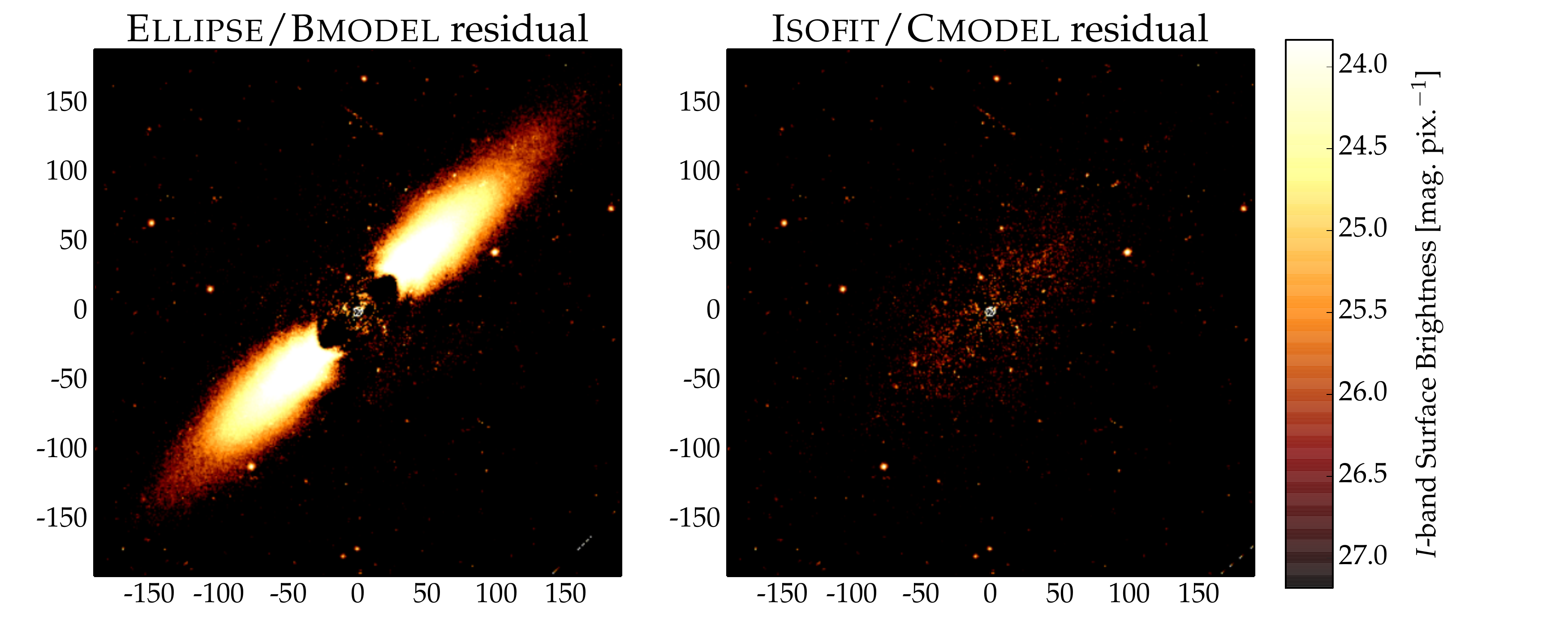}
	\caption{NGC 3610. The panels are arranged analogously to Figure \ref{fig:eso2543-49}. The isophote shapes require higher-order harmonics to properly capture the combination of both boxyness (`square' along the minor axis) and diskyness (`pointy' along the major axis due to the embedded, nearly edge-on disk) within the same isophote.}
	\label{fig:n3610}
\end{figure*}

As mentioned in Section \ref{sec:NewAngle}, one of the goals of this work is to investigate whether a galaxy's more complex morphological features are encoded in its isophotal structure, and can therefore be quantified through the Fourier coefficients. While the $B_4$ harmonic coefficient has been shown in the past to correlate with structural and physical properties of elliptical galaxies, the higher orders and late-type galaxies have been left quite unexplored (largely due to the improper expression of the isophote contours). With this new formalism it is now possible to explore this parameter space. 

If we observe the isophote contours of NGC 2549 in the upper panels (centre and right) of Figure \ref{fig:n2549}, we notice how the isophotes transition from boxy (inner) to disky (outer). The shapes at the transition are not elliptical however (even though $B_4 = 0$), but rather a combination of boxy \textit{and} disky (green contour), reminiscent of the $B_6$-corrected ellipse illustrated in Figure \ref{fig:coef246} (bottom right panel). The transition occurs precisely in the radial range where the X-shape/peanut feature is most obvious in the image. Since it is clear that higher-order harmonics are necessary to capture this morphological feature, the harmonic coefficients actually `measure' the \textit{peanutness} in this way. This is an important result: with  the degree of peanutness quantifiable this way, it is now possible to directly and quantitatively compare observations of real galaxies with simulations (such as e.g., \citealt{Athanassoula2005}, \citealt{Athanassoula2014}). 

A second galaxy, NGC 3610, which displays a combination of boxy \textit{and} disky isophotes was modelled in this section. This galaxy is consistent with having a thick boxy bulge, i.e., the extent of the disk component is less than or comparable to that of the bulge. In this work the terminology `boxy/peanut (B/P) bulge' is avoided, however, precisely because boxyness is expressed through $B_4$, whereas the isophotes of these structures are associated with different/additional harmonics. In stead, these structures are referred to as X-shaped/peanut (X/P) bulges. The data was again retrieved from the HLA, and consisted of an $I$-band image taken with WFPC2 (filter F814W; PI: B. Whitmore). The resultind models and residual maps corresponding to this galaxy are displayed in Figure \ref{fig:n3610}. Once again, standard \elli\; can not model the higher harmonic orders, and the residual image is dominated by an artificial streak along the disk plane. 

\begin{figure*}
	\centering
	\includegraphics[width=1.\columnwidth]{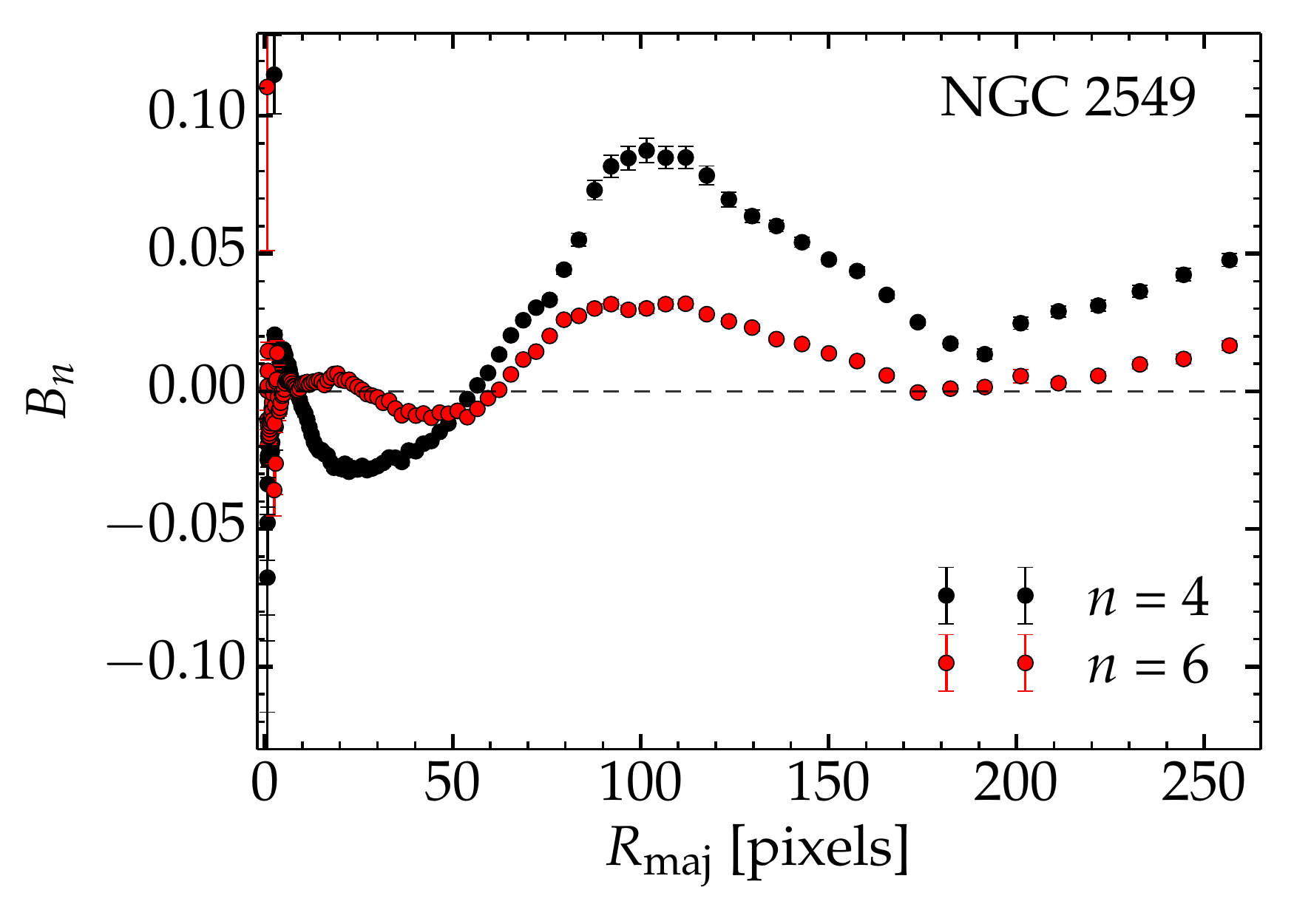}
	\includegraphics[width=1.\columnwidth]{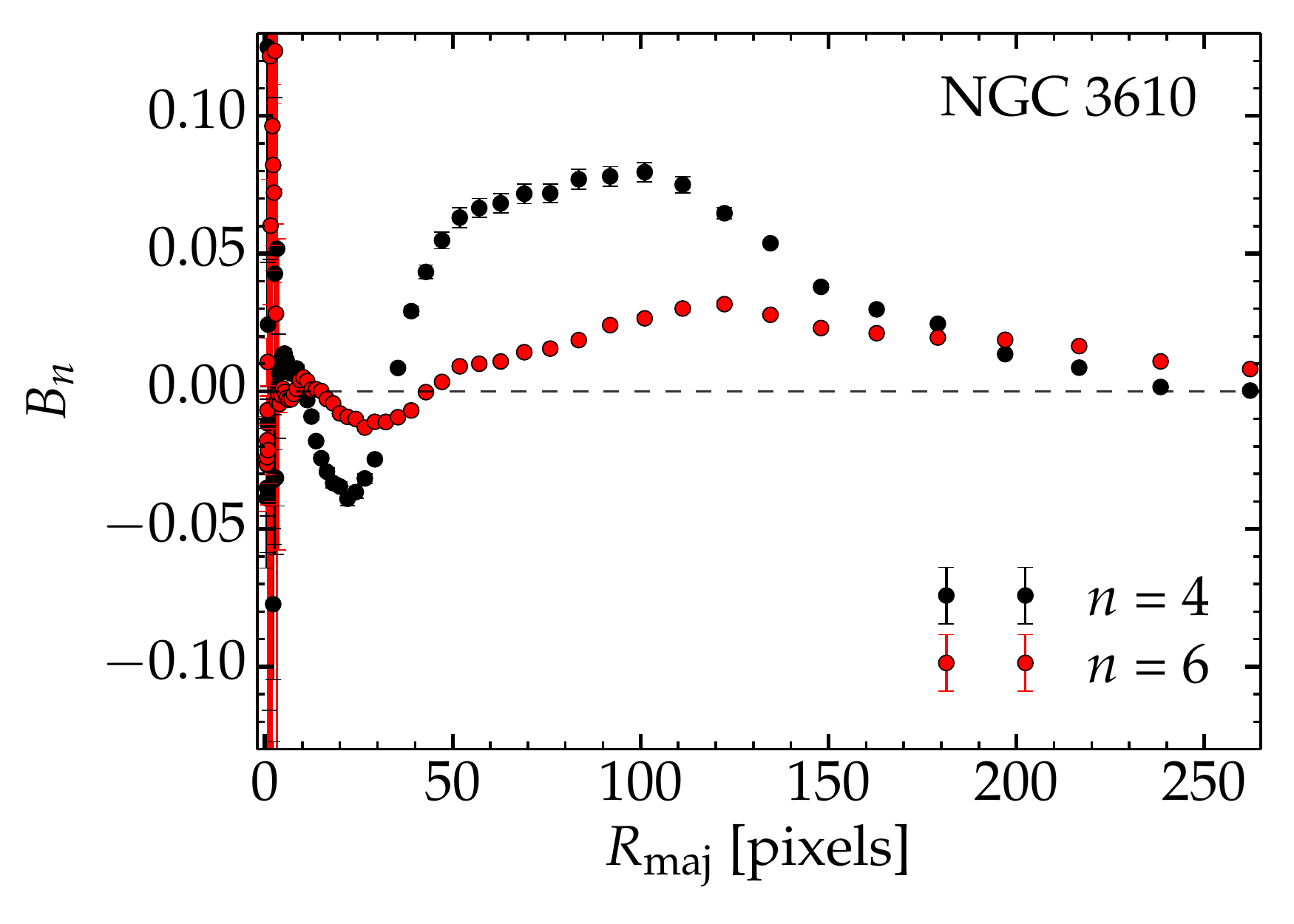}
	\caption{The fourth and sixth Fourier coefficient profiles corresponding to the isophotes of NGC 2549 (left panel) and NGC 3610 (right panel), normalised by the local semi-major axis and intensity gradient (see Equation \ref{equ:normalisation}) and plotted as a function of semi-major axis length. In NGC 2549 there is a transition between boxy (negative $B_4$ at $ R_{maj} \lesssim$ 50 pixels) and disky  (positive $B_4$ at $ R_{maj} \gtrsim$ 50 pixels) isophotal shapes, and the higher order $B_6$ term is required to capture the `peanut' feature. NGC 3610 has a similar transition in $B_{4}$ and also requires a $B_6$ contribution.}
	\label{fig:bn_coeff}
\end{figure*}

Figure \ref{fig:bn_coeff} displays the $B_4$ and $B_6$ coefficients of NGC 2549 and NGC 3610, computed with \ifit, as they change with radius (along the major axis). The coefficients are re-normalised by the local gradient and local semi-major axis length (Equation \ref{equ:normalisation}), such that 

\begin{equation}
B_n \rightarrow- B_n \left( a \frac{\partial I}{\partial a} \right)^{-1} ,
\label{equ:normalisation}
\end{equation}

where $a$ is the local semi-major axis length of the ellipse. This is a common way of expressing the coefficients as (dimensionless) deviations from ellipticity, as opposed to concrete intensity corrections. The transition between boxy and disky isophotes ($B_4$ crossing 0) occurs in both galaxies, and also $B_6$ has non-zero amplitude for most of the radial range, which indicates that it plays a part in capturing the isophote shape. Such accurate information was not available with the old formalism used in \textsc{IRAF}.\\

\subsection{Photometry of Galaxy Substructure}\label{sec:substruct}

Perhaps the most basic use of a high-quality model of a galaxy is the study of residual substructure, such as star clusters, globular clusters (GCs), satellites, streams etc. Once the model is subtracted from the original image, all of this substructure is left behind in the residual image. We have seen in Section \ref{sec:NA-TestCase} that \cmo\; makes it possible to perform photometry on residual images quite straightforwardly with tools such as the Aperture Photometry Tool. In this section we qualitatively explore this further, and extend the study to modelling galaxies with more complex isophotal structures such as misaligned bars. To begin with, the interesting barred lenticular (SB0) galaxy NGC 936 was the first candidate to be modelled. Due to this galaxy's orientation (slightly inclined but close to face-on view) and components (bulge, disk and \textit{barlens}; see \citealt{Laurikainen_barlens2011}), its isophote structure is relatively complicated. The data was retrieved from the publicly available archive of the Spitzer Survey of Stellar Structure in Galaxies (S4G\footnote{http://irsa.ipac.caltech.edu/data/SPITZER/S4G/}), and consisted of an infrared (3.6$\mu$) Spitzer observation (IRAC 1 instrument, PI: K. Sheth). Figure \ref{fig:n936} displays the results: the isophote shape starts off close to elliptical in the central regions, then it transitions to lemon-shaped and parallelogram-shaped (when capturing the barlens) and finally returns to a shape close to elliptical, when the barlens no longer dominates and the disk takes over. The isophotes also rotate in position angle with increasing radius, a process which is again driven by the barlens. Because of all these strong deviations from ellipticity, the 2D model built with \bmo\; fails to reproduce the light distribution in this galaxy and after its subtraction from the image, a residual map remains that is heavily contaminated by twisting features, both positive and negative. The much more realistic 2D model built with \cmo\; (top right) and the associate residual map (bottom right) demonstrate that the formalism proposed here is not limited to edge-on systems, but can competently handle any type of galaxy. With such good quality results, it is now possible to make stronger claims about galaxy morphology and also gain better insight into the substructure than before. The previous contaminated residual images made it difficult to analyse globular clusters because the artificial features often hid them in oversubtracted regions or covered them through the bright artificial streaks. This not being a problem with our new approach, it is now possible to perform GC number statistics (e.g., GC number density radial profiles) as well as more accurate photometry.

The second galaxy modelled in this section was deliberately chosen from a survey dedicated to GC studies, namely the SLUGGS\footnote{http://sluggs.swin.edu.au/} survey (\citealt{Usher2012}, \citealt{Brodie2012}). The galaxy in question, NGC 4111, is another example of a disk viewed at close to 90$\degree$ inclination, and is in fact the `flattest' galaxy modelled in this paper. The data was again retrieved from S4G archive, and consisted of an infrared (4.5$\mu$) Spitzer observation (IRAC 2 instrument, PI: K. Sheth). The results are displayed in Figure \ref{fig:n4111}, where again we notice the contaminated residual image produced with \bmo\;as well as the `cleaner' residual obtained with with \cmo.\\

\begin{figure*}
	\centering
	\includegraphics[width=1.\textwidth]{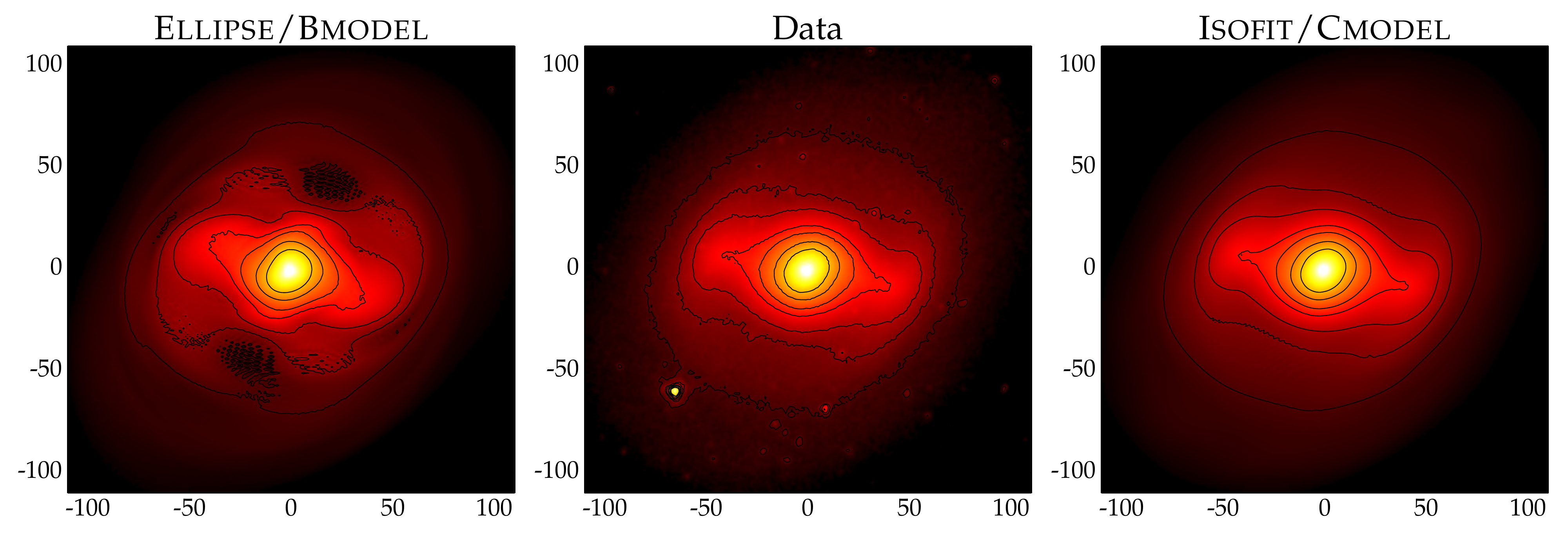}\\
	\flushright
	\includegraphics[width=.9\textwidth]{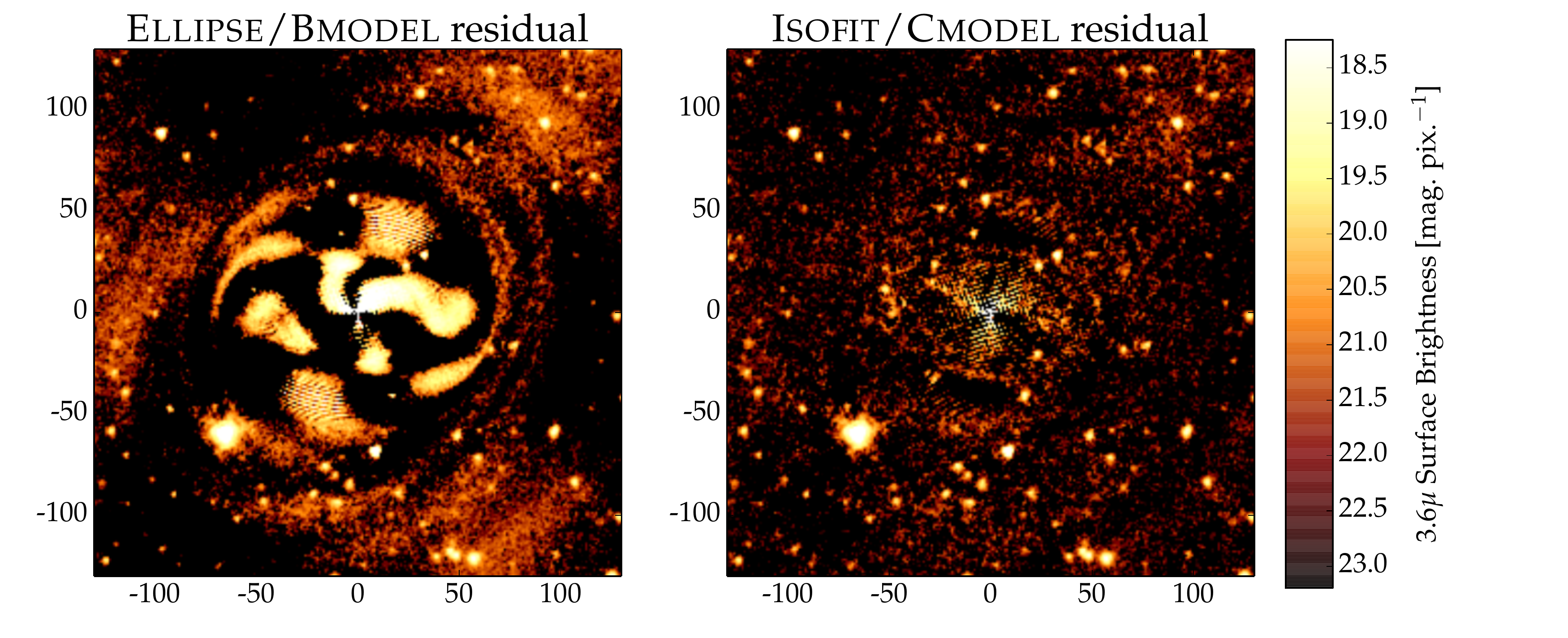}
	\caption{NGC 936: an SB0 galaxy with a barlens and complex isophotes. The panels are arranged analogously to Figure \ref{fig:eso2543-49}. The barlens induces strong departures from pure elliptical shapes, as well as substatial (close to $ 90 \degree$) isophote rotation.}
	\label{fig:n936}
\end{figure*}

\section{Summary and Outlook}\label{sec:sum}

This paper proposes a new formalism in which the isophotes of galaxies are expressed as a function of an angular parameter more natural to ellipses, namely the `eccentric anomaly'. This replaces use of the polar angle and significantly improves on past algorithms, such as the \textsc{IRAF} tasks \elli\;and \bmo. Isohotes with significant deviations from pure ellipses can now be accurately modelled, opening the door to new science.

This new formalism is implemented in the tasks \elli\;and \bmo, which are renamed \ifit\;and \cmo, respectively, in order to differentiate the new implementation from the old standard. The method is tested on an edge-on galaxy, ESO 243-49, and is found to bring considerable improvements in both 1D (major axis surface brightness profile) and 2D (model and residual image). Specifically, with \ifit\;the true surface brightness profile of ESO~243-49 is recovered, in shape and normalisation, while with \elli\;it was systematically underestimated, the discrepancy reaching a maximum of 0.77 mag arcsec$^{-2}$ (a factor of $\sim$ 2 in brightness) at $R_{maj} = 16.5$ arcsec for the galaxy in question. In addition, the second new task \cmo\;was found to produce a realistic 2D representation of the light distribution, which, after being subtracted from the image, left behind a high quality residual image on which photometry was possible. The $H$-band magnitude of the optical counterpart of HLX-1 was measured to be $m_H = 22.68 \pm 0.27$ mag.

\begin{figure*}[!ht]
	\centering
	\includegraphics[width=1.\textwidth]{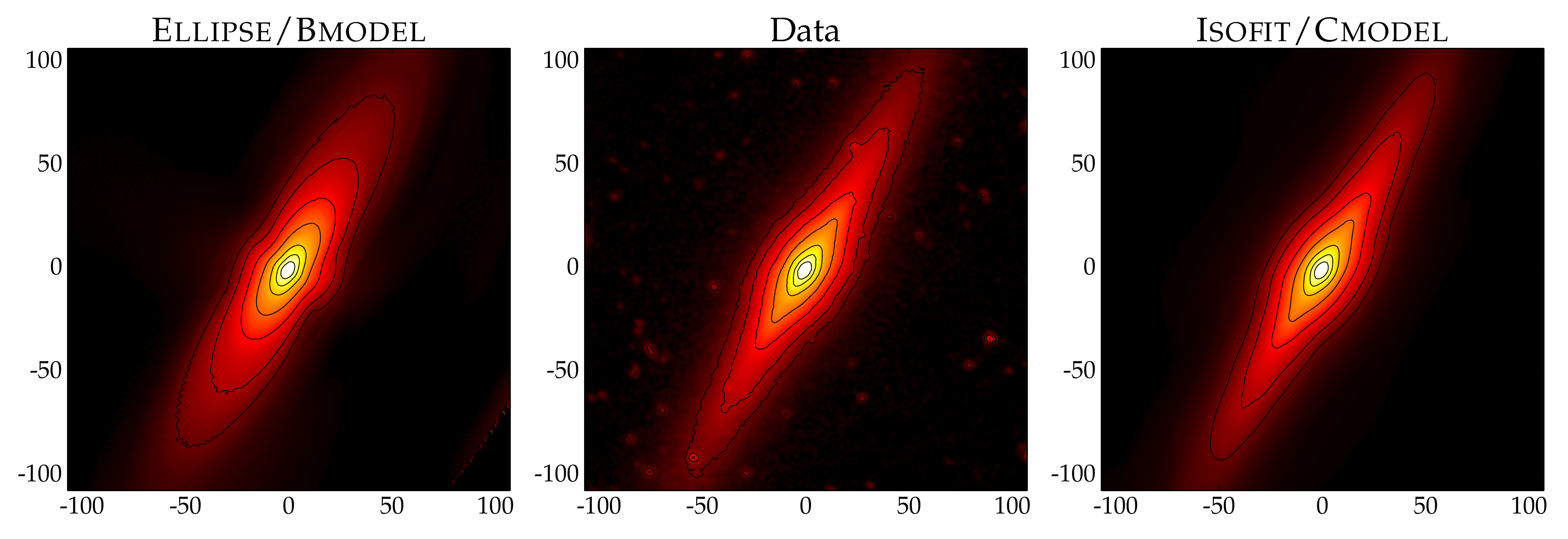}\\
	\flushright
	\includegraphics[width=.9\textwidth]{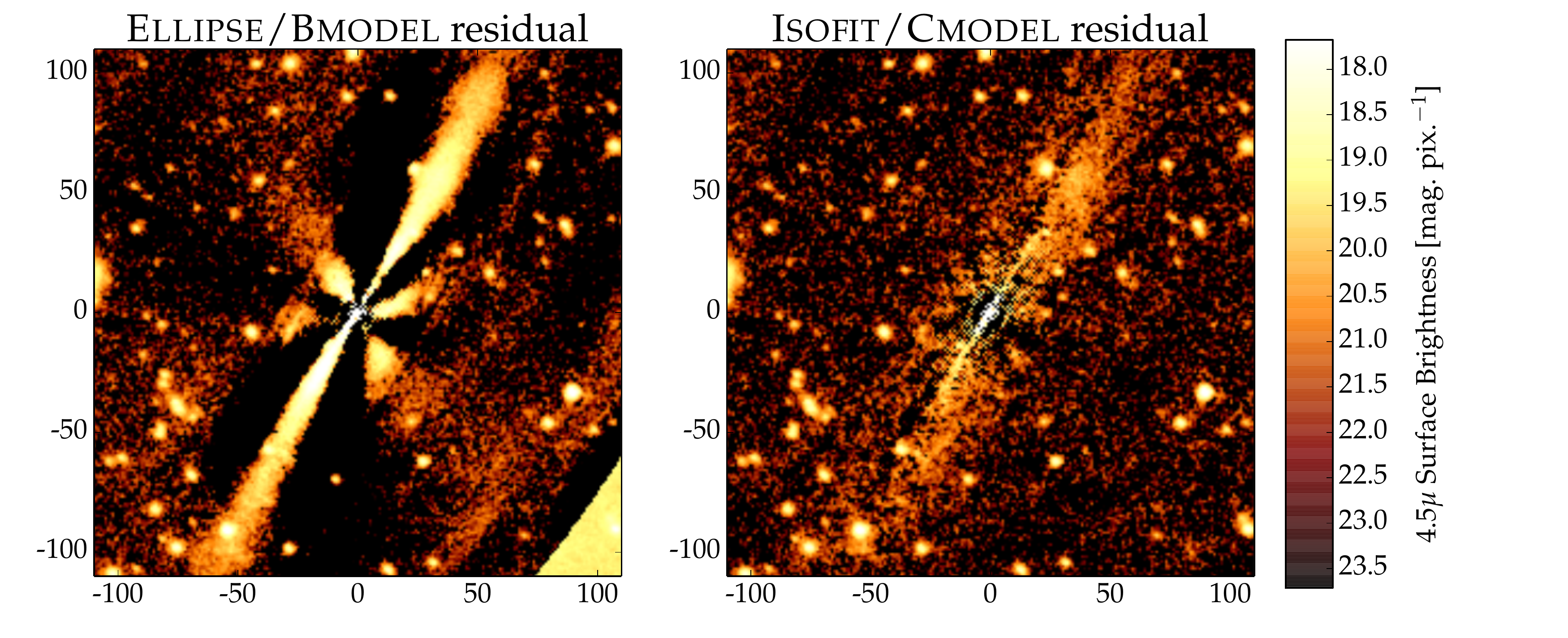}
	\caption{NGC 4111: an edge-on galaxy surrounded by globular clusters. The panels are arranged analogously to Figure \ref{fig:eso2543-49}.  The artificial cross in the residual obtained from \bmo\; (bottom left) makes it hard to detect all the globular clusters, and contaminates the computation of their brightness. Application of the new \cmo\; obtains a much better residual image (bottom right).}
	\label{fig:n4111}
\end{figure*}

This work makes it possible to study a wide range of photometrically interesting objects, and two particular scienctific applications (beyond galaxies with near edge-on disks) were further considered: galaxies with X-shaped/peanut bulges, and photometry of galaxy substructure. By expressing isophotes with a more natural metric, the higher order Fourier moments now carry meaning and can potentially quantify morphological features such as the peanut feature in galaxy bulges. This conclusion, however, is only based on the two galaxies modelled in the paper and an exhaustive and more quantitative study on a larger sample is reserved for future work to confirm this hypothesis. While the full power of this new formalism is most obvious in edge-on galaxies, it is not limited to these systems, and in fact is fully applicable to any galaxy. The task \ifit\;is capable of capturing complex isophote shapes with high deviations from ellipticity and, in the limit of low ellipticity or low deviations from ellipticity, the results of \elli\;are fully recovered.

The case-studies discussed in the above sections represent a small sample of the much wider range of science made possible by the better mathematical description of isophotal structure used in \ifit.
One such application involves the modelling of compact, low-mass early type galaxies. Current investigations on such galaxies (e.g., \citealt{Guerou_etal2015}) still model them with \elli\;and thus fail to capture interesting features such as embedded disks, which in turn leads to the familiar cross-pattern in residual maps. Because of this, often times the harmonics are deliberately not included, and the presence of disks/diskyness is pointed out by the cross pattern in the residual image.

Additionally, recent studies have shown that the bulge of the Milky Way has an X/peanut-shaped structure (\citealt{Wegg2015}, \citealt{DiMatteo2015}, \citealt{Qin2015}). Since this makes it the nearest X/P bulge in an edge-on galaxy, it is a prime candidate to study with \ifit\;and investigate potential connections between the X/P characteristic and $n > 4$ Fourier coefficients.

Finally, this method can be applied to the interesting shell-galaxies (e.g., \citealt{Duncan_shells}, \citealt{delBurgo2008}). These are elliptical galaxies marked with shell-like or ring-like structures, thought to be the remnants of past merger events.

All of these constitute extensive projects in their own right and are deferred to future work.\\

\section{Acknowledgements}

The author expresses his sincere gratitude to Alister W. Graham, for his expert advice, insightful and stimulating discussions, continual encouragement and for a careful reading of key parts of this manuscript. The author expresses thanks to Luke Hodkinson, who provided helpful technical support for modifying the source code in the \textsc{IRAF} package \iso\;to implement the formalism introduced in this work.

\clearpage

\begin{appendix}
\setcounter{figure}{0}
\renewcommand{\thefigure}{A\arabic{figure}}

\setcounter{equation}{0}
\renewcommand{\theequation}{A.\arabic{equation}}

\newcommand{\area}{\mathcal{A}}
\newcommand{\dif}{\textrm{d}}

The formalism proposed in this work is fully consistent with re-mapping the major-axis surface brightness profile $\mu(R_{\rm maj})$, computed by \ifit, onto the `equivalent axis' $R_{\rm eq}$. The `equivalent radius' $R_{\rm eq}$ is the radius of the `circularised' isophote of semi-major axis length $R_{\rm maj}$ such that circle encloses an equal area as that of the isophote. This transformation is routinely performed on $\mu(R_{\rm maj})$ following a bulge/disk decomposition of $\mu(R_{\rm maj})$, because it allows the use of circular symmetry in computing the integrated surface brightness (i.e., the magnitude) of the model components. The deviations from a purely elliptical shape brought by the \ifit\;formalism {\it conserve} the enclosed area, as is demonstrated below. \\

{\it I.)} Consider a surface enclosed in a circle of radius $a$. In plane-polar co-ordinates, the surface element  d$S$ is given by:

\begin{equation}
\dif S = r\dif r\dif \phi .
\end{equation}

The total area $\area$, obtained by integrating d$S$ in the range $r \in [0,a]$ and $\phi \in [0,2\pi]$ is the usual

\begin{equation}
 \area = {\displaystyle \int_{0}^{a} \int_{0}^{2\pi}} r\;\dif r \dif \phi  = \pi a^2 . 
\end{equation}

{\it II.)} Consider re-scaling the circle along the $y$-axis by a factor $b/a$. The resulting shape is an ellipse with semi-major axis $a$ and semi-minor axis $b$ (Figure \ref{fig:apx}, panel a). Moreover, the area $\area$ is also re-scaled by the same factor: $\area_{ell} = \area \ b/a = \pi a^2 (b/a) = \pi ab$,  the usual expression for the area of an ellipse. Further, a point $F$ on the circle (defined py a central angle $\phi$) becomes the point $F'$ on the ellipse, and is now defined by a central angle $\psi$. The two angles are related to each other through the axis ratio $b/a$ or ellipticity $e=1-a/b$, through Equation \ref{equ:conversion}. This is in fact the geometric definition of the `eccentric anomaly' $\psi$.\\

{\it III.)} Consider again the circular surface, but perturbed with a set of sinusoidal waves, as a function of polar angle (e.g., Figure \ref{fig:apx}, panel b), such that

\begin{equation}\label{equa:harm}
r(\phi) = r [1 + A_n \textrm{sin}(n\phi) + B_n \textrm{cos}(n\phi)],
\end{equation}

where $A_n, B_n$ are constant coefficients and $n$ is an integer. The new surface element (denoted as d$S'$) becomes:

\begin{equation}
\dif S' =  r [1 + A_n \textrm{sin}(n\phi) + B_n \textrm{cos}(n\phi)]\dif r \dif\phi .
\end{equation}

As before, the total area of the perturbed shape (denoted as $\area '$) is given by integrating the surface element in the same limits:

\begin{equation}
\begin{split}
 \area' &= {\displaystyle \int_{0}^{a} \int_{0}^{2\pi}} r [1 + A_n \textrm{sin}(n\phi) + B_n \textrm{cos}(n\phi)]\;\dif r \dif \phi  \\ 
         &= {\displaystyle\int_{0}^{a}\int_{0}^{2\pi}} r\;\dif r \dif \phi + A_n {\displaystyle \int_{0}^{a}\int_{0}^{2\pi}} r \textrm{sin}(n\phi)\;\dif r \dif \phi + B_n {\displaystyle \int_{0}^{a}\int_{0}^{2\pi}} r \textrm{cos}(n\phi)\;\dif r \dif \phi \\
         &= \pi a^2 + \frac{a^2}{2} {\displaystyle \int_{0}^{2\pi}} \textrm{sin}(n\phi) \dif \phi  + \frac{a^2}{2} \int_{0}^{2\pi} \textrm{cos}(n\phi) \dif \phi \\
         &= \pi a^2 + \frac{a^2}{2n} [-1 + 1  + 0 - 0] \\
         &= \pi a^2 \equiv \area, \forall\; n
\end{split}
\end{equation}

\begin{figure}[!h]
\centering
	\includegraphics[width=1.\textwidth]{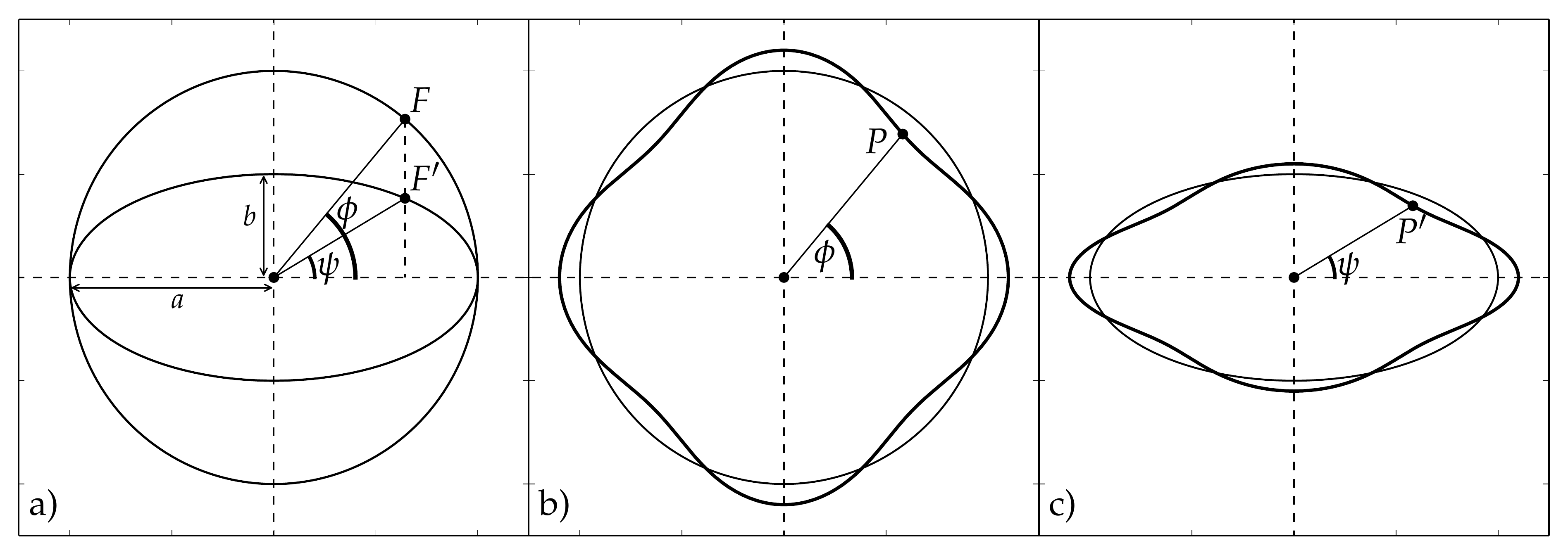}
	\label{fig:apx}
	\caption{{\it Panel} a): When the circle of radius $a$ is `squeezed' along the $y-$axis by a factor $b/a$, it results into an ellipse of semi-minor axis $b$ and semi-major axis $a$. The point $F$ on the circle (defined by $\phi$) corresponds to $F'$ on the ellipse, which is described by the `eccentric anomaly $\psi$. {\it Panel} b): The circle in panel a) distorted through Equation \ref{equa:harm}, with $n=4$, $A_n=0$, $B_n=0.1$ (thick line). A point $P$ along the circumference is defined by the polar angle $\phi$. {\it Panel} c): When the distorted circle in panel b) is `squeezed' into a distorted ellipse, the same point (now denoted by $P'$) is now defined by the `eccentric anomaly' $\psi$. }

\end{figure}

The sinusoidal functions change the area as a function of $\phi$ but the net change is zero over the full range [0,2$\pi$], i.e., the positive `bumps' are exactly cancelled out by the negative `dips' over the whole azimuthal range, independently of the choice of $n$. \\

{\it IV.)} Consider now the perturbed shape `squeezed' along the $y$ direction by a factor $b/a$ (Figure \ref{fig:apx}, panels b and c), in the same way as the circular shape was re-scaled (panel a). A point $P$ on the perturbed circle, which was described by the polar angle $\phi$ (panel b), becomes the point $P'$ on the perturbed ellipse (panel c), described by the eccentric anomaly $\psi$ as before. This is at the heart of the the new formalism - a Fourier wave on a circle needs a different co-ordinate system on the corresponding ellipse. As a consequence of the re-scaling (`squeezing'), all areas (in panel b) are reduced by the same factor ($b/a$) in panel c. The area inside and outside of the ellipse is thus reduced by $b/a$, but there is still, obviously, an equal amount of area inside and outside. The area of the ellipse is therefore conserved following the perturbation.\\

{\it V.)} From the output of \ifit\; ($\left< I_{ell} \right>, R_{\rm maj}$ and $A_n, B_n$ -- see Section \ref{sec:Iso-Fit} for the explanation) the major axis surface brightness profile $\mu(R_{maj})$ can be constructed in 2 equivalent ways:\\

\begin{enumerate}
\item By applying the perturbations to the intensity/surface brightness ($I_{\psi=0} \neq \left< I_{ell} \right>$) -- keeping the isophote shape elliptical and distorting the intensity distribution along its circumference. This is how \ifit\;works.

\item By keeping $\left< I_{ell} \right>$ fixed and applying the corrections to $R_{\rm maj}$ -- distorting the ellipse shape and keeping the intensity fixed. This step additionally requires the intensity gradient, which is also provided by \ifit\; (\elli\; also computes all these quantities, though with incorrect $A_n, B_n$), to re-normalise the coefficients (such that they correspond to $R_{\rm maj}$ perturbations).
\end{enumerate}

Performing the distortion the second way is analogous to case {\it IV.)}: a given isophote is a distorted ellipse whose area is the same as that of the original, pure ellipse ($\area_{iso} = \area_{ell} = \pi ab = \pi a^{2} (1-e) \equiv \pi R_{\rm maj}^{2} (1-e)$). The intensity associated with this isophote area is $\left< I_{ell} \right>$ (which is converted into the surface brightness $\mu_{ell}$). Therefore, if the area is expressed as a circular area, its radius $R_{\rm eq}$ can be obtained as:

\begin{equation}
\pi R_{\rm eq}^{2} = \pi R_{\rm maj}^2 (1-e)
\end{equation}

\begin{equation}
R_{\rm eq} = R_{\rm maj} \sqrt{1-e}
\end{equation}

With the surface brightness profile now expressed along the equivalent axis, it is straightforward to compute magnitudes in the usual fashion.\\

\end{appendix}

\bibliography{references}

\end{document}